\documentclass[aps,showpacs,twocolumn]{revtex4}

\usepackage{graphicx}
\usepackage{amsmath}

\def\ov#1{\overline{#1}}

\def\wt#1{\widetilde{#1}}
\def\vb#1{\mbox{\boldmath$#1$}}
\def\pd#1#2{\frac{\partial #1}{\partial #2}}

\def\wh#1{\widehat{#1}}
\def\bdot{\,\vb{\cdot}\,}
\def\btimes{\,\vb{\times}\,}

\def\bhat{\wh{{\sf b}}}
\def\bstar{{\sf b}^{*}}

\def\exd{{\sf d}}

\def\L{\mathcal L}

\newcommand{\bc}{\begin{center}}
\newcommand{\ec}{\end{center}}
\newcommand{\bt}{\begin{tabbing}}
\newcommand{\et}{\end{tabbing}} 
\newcommand{\be}{\begin{eqnarray*}}
\newcommand{\ee}{\end{eqnarray*}}
\newcommand{\bs}{\begin{slide}}
\newcommand{\es}{\end{slide}}

\def\ec{\epsilon_c}
\def\em{\epsilon_m}
\def\epsm{\epsilon_m}
\def\betaip{{\beta_i'}}
\def\betaep{{\beta_e'}}
\def\betatotp{{\beta'_{\rm tot}}}
\def\kpl{k_\parallel}

\def\Alfven{Alfv\'{e}n}

\begin{document}

\title{Nonlinear finite-Larmor-radius effects in reduced fluid models}

\author{A.~J.~Brizard}
\affiliation{Department of Chemistry and Physics, Saint Michael's College, Colchester, VT 05439} 

\author{R.~E.~Denton and B.~Rogers} 
\affiliation{Department of Physics and Astronomy, Dartmouth College, Hanover, NH 03755} 

\author{W.~Lotko}
\affiliation{Thayer School of Engineering, Dartmouth College, Hanover, NH 03755} 

\begin{abstract}
The polarization and magnetization effects associated with the dynamical reduction leading to the nonlinear gyrokinetic Vlasov-Maxwell equations are shown to introduce nonlinear finite-Larmor-radius effects into a set of nonlinear reduced-fluid equations previously derived by Lagrangian variational method [A.~J.~Brizard, Phys.~Plasmas {\bf 12}, 092302 (2005)]. These intrinsically nonlinear FLR effects, which are associated with the transformation from guiding-center phase-space dynamics to gyrocenter phase-space dynamics, are different from the standard FLR corrections associated with the transformation from particle to guiding-center phase-space dynamics. We also present the linear dispersion relation and results from a nonlinear simulation code using these reduced-fluid equations.  The simulation results (in both straight and dipole geometries)  demonstrate that the equations describe the coupled dynamics of \Alfven\ and sound waves and that the total simulation energy is conserved.
\end{abstract}

\begin{flushright}
June 25, 2008
\end{flushright}


\maketitle

\section{Introduction}

The development of plasma fluid models that incorporate finite-Larmor-radius (FLR) corrections has a rich history in plasma physics (see Ref.~\cite{Schnack_etal} and references therein and Ref.~\cite{Hazeltine_Meiss}). Plasma fluid models offer the great advantage of simple tractability of a few fields with excellent small-scale resolution of space-time scales. One of the disadvantages of fluid models over kinetic models, however, is that specific features of the orbital particle dynamics are lost by averaging over particle momentum (or velocity) space. 

Various fluid models have, thus, been built by incorporating one or more of the following kinetic effects: standard FLR effects associated with the transformation from particle phase-space dynamics to guiding-center phase-space dynamics \cite{HW_1983,HHM_1986,Brizard_1992,SPH_1995,Belova_2001}; wave-particle resonances (e.g., Landau damping) \cite{SHD_1997,SH_2001}; and separate contributions from magnetically-trapped and untrapped particles in tokamak geometry \cite{BH_1996,SKW_2005}. The Hamiltonian structure of reduced magnetohydrodynamic equations \cite{MH_1984} has also led to the development of sophisticated reduced fluid models \cite{HHM_1986,HHM_1987} that retain a Hamiltonian structure as well as preserve important conservation laws. In particular, it is often stressed in the derivation of reduced fluid models that the energy-conservation property (important for accurate numerical implementation) represents an important property to be preserved through reduction. The existence of a Lagrangian variational principle from which reduced nonlinear fluid equations are derived guarantees the derivation of an exact energy conservation law by the Noether method.

The purpose of the present work is to (i) generalize the previous work by Brizard \cite{Brizard_2005}, (ii) provide a new interpretation for the reduced-fluid equations presented in Ref.~\cite{Brizard_2005} in terms of nonlinear FLR corrections, and (iii) demonstrate the properties of these equations with a linear dispersion relation (to uncover the types of waves they describe) as well as linear and nonlinear numerical simulations in straight and magnetic dipole geometries to investigate energy conservation. 

\subsection{Perturbed electric and magnetic fields}

First, the perturbed electric and magnetic fields
\begin{eqnarray}
{\bf E} & \equiv & -\,\nabla\Phi \;-\; \frac{\bhat_{0}}{c}\;\pd{A_{\|}}{t}, \label{eq:E_def} \\
{\bf B}_{\bot} & \equiv & \nabla \times (A_{\|}\;\bhat_{0})
\label{eq:B_def}
\end{eqnarray}
used in the present work are expressed in terms of the perturbed scalar potential $\Phi$ and the perturbed vector potential $A_{\|}\,\bhat_{0}$, where we assume that the perturbed magnetic field is perpendicular to the (time-independent) background magnetic field ${\bf B}_{0} \equiv B_{0}\,\bhat_{0}$. 

Note that the formula for the perturbed magnetic field used in Ref.~\cite{Brizard_2005} was not in general divergenceless (unless $\nabla\btimes\bhat_{0} = 0$, i.e., $\bhat_{0}$ is the gradient of a scalar), but Eq.~(\ref{eq:B_def}) is explicitly divergenceless. Because of this modification, some of the equations in this paper differ from those of Ref.~\citep{Brizard_2005}. In particular, while magnetic-curvature effects were absent from the parallel Amp\`{e}re equation in Ref.~\citep{Brizard_2005}, these effects are taken into account in the present work \cite{footnote_0}.  Note also that the condition $\bhat_{0}\bdot{\bf B}_{\bot} \equiv 0$ implies the assumption $\bhat_{0}\bdot\nabla\btimes\bhat_{0}=(\bhat_{0}/B_0)\bdot\nabla\btimes{\bf B}_{0}= 0$, i.e., we assume that no background current $J_{\|0} \propto \bhat_{0}\bdot\nabla\btimes{\bf B}_{0}$ flows along the background field lines (e.g., magnetic dipole geometry). This assumption is adopted here in order to simplify our discussion of nonlinear FLR effects and we note that the condition 
$\bhat_{0}\bdot\nabla\btimes{\bf B}_{0} \neq 0$ can easily be restored in our nonlinear reduced fluid equations (see Ref.~\cite{SSB} for example). 

\subsection{Standard and nonlinear FLR corrections}

Second, the new interpretation for the reduced-fluid equations of Ref.~\cite{Brizard_2005} is given in terms of nonlinear FLR corrections that arise from nonlinear gyrokinetic theory \cite{Brizard_Hahm}. Standard FLR corrections are associated with the transformation from particle phase space to guiding-center phase space. In particular, the gyroradius vector 
\begin{equation}
\vb{\rho}_{{\rm gc}} \;\equiv\; {\sf T}_{{\rm gc}}^{-1}{\bf x} \;-\; \ov{{\bf x}}
\label{eq:rho_gc}
\end{equation} 
is interpreted as the displacement between the representation of the particle position in guiding-center phase space ${\sf T}_{{\rm gc}}^{-1}{\bf x}$ and the guiding-center position $\ov{{\bf x}}$ (here, ${\sf T}_{{\rm gc}}^{-1}$ denotes the push-forward operator associated with the guiding-center phase-space transformation; see Appendix \ref{sec:push} for further details). 

Standard FLR corrections in plasma reduced-fluid formalism are based on the fact that a guiding-center particle feels electromagnetic fields that are averaged over a gyration period. Hence, a guiding-center particle (with mass $m$ and charge $q$) feels the gyroangle-averaged potential
\begin{equation}
\langle \Phi(\ov{{\bf x}} + \vb{\rho}_{0})\rangle \;=\; \Phi(\ov{{\bf x}}) \;+\; \frac{\mu B_{0}}{2m\Omega_{0}^{2}}\;
\ov{\nabla}_{\bot}^{2}\Phi(\ov{{\bf x}}) \;+\; \cdots,
\label{eq:FLR}
\end{equation}
where, by definition, the gyroangle-average of the lowest-order gyroradius $\vb{\rho}_{{\rm gc}} = \vb{\rho}_{0} + \cdots$ vanishes (i.e., $\langle
\vb{\rho}_{0}\rangle \equiv 0$) and the lowest-order nonvanishing FLR correction involves the gyroangle-averaged dyadic product $\langle\vb{\rho}_{0}\vb{\rho}_{0}\rangle \equiv \frac{1}{2}\,\rho_{0}^{2}\;{\bf I}_{\bot}$, where $\Omega_{0} = qB_{0}/mc$ denotes the (signed) gyrofrequency, $\rho_{0} = [2\mu B_{0}/(m\Omega_{0}^{2})]^{1/2}$, and ${\bf I}_{\bot} = {\bf I} - \bhat_{0}\bhat_{0}$. The ordering of the standard FLR correction in Eq.~(\ref{eq:FLR}) involves the factor $\lambda \equiv k_{\bot}^{2}\rho_{0}^{2}$, where $k_{\bot} = |{\bf k}_{\bot}|$ denotes the perpendicular wavenumber. In Fourier ${\bf k}$-space (where $\nabla_{\bot} \rightarrow i\,{\bf k}_{\bot}$), the expansion (\ref{eq:FLR}) may be summed up as $\langle\Phi\rangle \rightarrow 
J_{0}(\lambda)\, \Phi_{{\bf k}}$, where $J_{0}(\lambda)$ is the zeroth-order Bessel function \cite{Brizard_1992,SPH_1995,Belova_2001}.

In nonlinear gyrokinetic theory \cite{Brizard_Hahm}, the asymptotic elimination of the gyroangle dependence reintroduced by the perturbation of guiding-center dynamics by low-frequency electromagnetic fluctuations is carried by the transformation from guiding-center phase space to gyrocenter phase space. The combination of the guiding-center and gyrocenter phase-space transformations introduces the gyrocenter displacement
\begin{equation}
\vb{\rho}_{{\rm gy}} \;\equiv\; {\sf T}_{{\rm gy}}^{-1}\left({\sf T}_{{\rm gc}}^{-1}{\bf x}\right) \;-\; {\sf T}_{{\rm gc}}^{-1}\ov{\ov{{\bf x}}},
\label{eq:rho_gy}
\end{equation} 
where ${\sf T}_{{\rm gy}}^{-1}({\sf T}_{{\rm gc}}^{-1}{\bf x})$ denotes the representation of the particle position in gyrocenter phase space and 
$\ov{\ov{{\bf x}}}$ denotes the gyrocenter position (here, ${\sf T}_{{\rm gy}}^{-1}$ denotes the push-forward operator associated with the gyrocenter phase-space transformation). The difference between the guiding-center position $\ov{{\bf x}}$ and the gyrocenter position $\ov{\ov{{\bf x}}}$ is directly proportional to the perturbed electromagnetic fields (\ref{eq:E_def})-(\ref{eq:B_def}) [see Eq.~(\ref{eq:rho_1}) below]. The lowest-order term in the gyrocenter displacement vector $\vb{\rho}_{{\rm gy}} = \vb{\rho}_{1} + \cdots$ is expressed in terms of the gyrocenter phase-space coordinates $(\ov{\ov{{\bf x}}}, v_{\|}, \mu, \zeta)$ as \cite{Brizard_Hahm}
\begin{eqnarray}
\vb{\rho}_{1} & = & -\;\{ S_{1},\; \ov{\ov{{\bf x}}} + \vb{\rho}_{0} \}_{{\rm gc}} \nonumber \\
 & = & -\;\frac{q}{mc}\;\left( \pd{S_{1}}{\zeta}\,\pd{\vb{\rho}_{0}}{\mu} \;-\; \pd{S_{1}}{\mu}\,\pd{\vb{\rho}_{0}}{\zeta} \right) \nonumber \\
 &  &+\; \frac{\bhat_{0}}{m}\;\pd{S_{1}}{v_{\|}} \;+\; \frac{c\bhat_{0}}{qB_{0}}\btimes\ov{\ov{\nabla}} S_{1}.
\label{eq:rho_gyro}
\end{eqnarray}
Here, $\{\;,\;\}_{{\rm gc}}$ denotes the guiding-center Poisson bracket (higher-order terms involving spatial gradients of the gyroradius 
$\vb{\rho}_{0}$ have been omitted), and the gyrocenter scalar field is given by the first-order FLR expression
\begin{equation}
S_{1} \;=\; \frac{q}{\Omega_{0}}\;\pd{\vb{\rho}_{0}}{\zeta}\bdot \left( {\bf E}_{\bot} \;+\; 
\frac{v_{\|}}{c}\;\bhat_{0}\btimes{\bf B}_{\bot} \right) \;+\; \cdots,
\label{eq:S1_def}
\end{equation}
where ${\bf E}_{\bot} = -\,\nabla_{\bot}\Phi$ and higher-order (standard) FLR terms have been omitted. The gyroangle-average of the gyrocenter displacement vector (\ref{eq:rho_gyro}) yields
\begin{eqnarray}
\langle\vb{\rho}_{1}\rangle & = & -\;\left\langle \{ S_{1},\; \vb{\rho}_{0} \}_{{\rm gc}}\right\rangle \;=\; -\;\frac{q}{mc}\;
\pd{}{\mu} \left\langle \vb{\rho}_{0}\,\pd{S_{1}}{\zeta} \right\rangle \nonumber \\
 & = & \frac{c}{B_{0}\Omega_{0}} \left( {\bf E}_{\bot} \;+\; \frac{v_{\|}}{c}\;\bhat_{0}\btimes{\bf B}_{\bot} \right),
\label{eq:rho_1}
\end{eqnarray}
where the guiding-center contribution $(\vb{\rho}_{0})$ and the gyrocenter contribution $(S_{1})$ combine to yield a nonvanishing gyroangle-averaged gyrocenter displacement.

The addition of the gyrocenter displacement vector (\ref{eq:rho_gyro}) to the standard gyroradius vector $\vb{\rho}_{0}$ thus means that the gyrocenter particle now experiences an averaged potential 
\begin{eqnarray}
\langle \Phi(\ov{\ov{{\bf x}}} + \vb{\rho}_{0} + \vb{\rho}_{1})\rangle & = & \Phi(\ov{\ov{{\bf x}}}) \;+\; \langle\vb{\rho}_{1}\rangle\bdot\ov{\ov{\nabla}}\Phi(\ov{\ov{{\bf x}}}) \nonumber \\ 
 &  &+\; \frac{\mu B_{0}}{2m\Omega_{0}^{2}}\;\ov{\ov{\nabla}}_{\bot}^{\;2}\Phi(\ov{\ov{{\bf x}}}) \;+\; \cdots,
\label{eq:FLR_nonlinear}
\end{eqnarray}
that combines the standard (guiding-center) and nonlinear (gyrocenter) FLR corrections. Note that the relative importance of nonlinear FLR effects relative to the standard FLR effects is represented by the ratio $q\Phi/T$ and, thus, nonlinear FLR effects dominate in the cold-ion-fluid limit when $q\Phi/T \gg 1$ \cite{DentonEA07,footnote_1}, i.e., when the linear $E \times B$ velocity (normalized to the thermal speed $v_{{\rm th}} = \sqrt{T/m}$) is large enough so that it satisfies the condition $|{\bf u}_{E}|/v_{{\rm th}} \gg k_{\bot}\rho_{{\rm th}}$. In order to focus our attention on the new nonlinear FLR corrections that appear in the reduced fluid equations of Ref.~\cite{Brizard_2005}, however, we set $k_{\bot}\rho_{{\rm th}} \equiv 0$ in the present work and postpone our discussion of standard FLR corrections in reduced fluid models to future work (see Ref.~\cite{SSB} for standard FLR corrections to a reduced electrostatic fluid model).

The combination of the guiding-center and gyrocenter phase-space transformations gives the relation ${\sf T}_{{\rm gy}}^{-1}\left({\sf T}_{{\rm gc}}^{-1}
{\bf x}\right) = \ov{\ov{{\bf x}}} + \vb{\rho}_{0} + \vb{\rho}_{1}$ between the particle position ${\bf x}$ and the gyrocenter position 
$\ov{\ov{{\bf x}}}$. Since the gyroangle-average of the gyrocenter-particle displacement 
\begin{equation}
\left\langle {\sf T}_{{\rm gy}}^{-1}\left({\sf T}_{{\rm gc}}^{-1}{\bf x}\right) \;-\; \ov{\ov{{\bf x}}}\frac{}{} \right\rangle \;=\; 
\langle\vb{\rho}_{1}\rangle \;\neq\; 0
\label{eq:rho1_average}
\end{equation}
does not vanish, it leads to the well-known polarization and magnetization effects in the gyrokinetic Maxwell's equations \cite{Brizard_Hahm} (see Appendix \ref{sec:push} for details concerning the dynamical reduction of the Vlasov equation by Lie-transform method \cite{Brizard_2006}).

By averaging the gyrocenter displacement vector (\ref{eq:rho_1}) over the gyrocenter Vlasov distribution in gyrocenter momentum space, we obtain the reduced-fluid displacement vector
\begin{eqnarray}
\vb{\rho}_{\bot} & \equiv & \frac{c}{B_{0}\Omega_{0}} \left( {\bf E}_{\bot} \;+\; \frac{u_{\|}}{c}\;\bhat_{0}\btimes
{\bf B}_{\bot} \right) \nonumber \\
 & = & \frac{\bhat_{0}}{\Omega_{0}}\btimes\left( {\bf u}_{E} \;+\; u_{\|}\;\frac{{\bf B}_{\bot}}{B_{0}} \right)
\label{eq:df_dipole}
\end{eqnarray}
where $u_{\|}$ denotes the gyrocenter-fluid parallel velocity, $u_{\|}\;{\bf B}_{\bot}/B_{0}$ represents the magnetic flutter velocity, and the linear perturbed $E\times B$ velocity is
\begin{equation}
{\bf u}_{E} \;\equiv\; \frac{c\bhat_{0}}{B_{0}}\btimes\nabla_{\bot}\Phi \;=\; {\bf E}_{\bot}\btimes
\frac{c\bhat_{0}}{B_{0}}. 
\label{eq:ExB_lin}
\end{equation} 
Using Eq.~(\ref{eq:df_dipole}), we define the effective potentials
\begin{equation}
\left. \begin{array}{rcl}
    \Phi_{\rho} & \equiv & \Phi \;-\; \vb{\rho}_{\bot}\bdot{\bf E}_{\bot} \\
 &  & \\
    A_{\|\rho}  & \equiv & A_{\|} \;-\; \bhat_{0}\bdot\vb{\rho}_{\bot}\btimes{\bf B}_{\bot}
\end{array} \right\},
 \label{eq:PhiA_rho}
\end{equation}
which include nonlinear finite-Larmor-radius (NFLR) corrections that are quite distinct from the standard FLR corrections; see Appendix 
\ref{sec:pol_disp} for further details concerning a physical interpretation of the reduced-fluid displacement (\ref{eq:df_dipole}) as well as a Lie-transform derivation of the effective potentials (\ref{eq:PhiA_rho}). The purpose of the present work is to investigate how polarization and magnetization effects manifest themselves in reduced fluid equations self-consistently derived from a Lagrangian variational principle.

The reduced gyrocenter-fluid moments $(n, {\bf u}, p_{\bot}, p_{\|})$ used in the present work, obtained as moments of the reduced Vlasov equation 
\cite{Brizard_2005,Brizard_2006}, are also expressed in terms of the physical (phys) fluid moments and the reduced-fluid displacement 
(\ref{eq:df_dipole}). According to Appendix \ref{sec:push}, the reduced-fluid density and parallel velocity $u_{\|} = \bhat_{0}\bdot{\bf u}$ are
\begin{equation}
\left. \begin{array}{rcl}
n_{{\rm phys}} & = & n \;-\; \nabla\bdot(n\,\vb{\rho}_{\bot}) + \cdots \\
 &  & \\
u_{\|{\rm phys}} & = & u_{\|} \;-\; \vb{\rho}_{\bot}\bdot\nabla u_{\|} + \cdots \\
\end{array} \right\},
\label{eq:red_phys}
\end{equation}
where higher-order nonlinear FLR effects are ignored. The reduced-fluid perpendicular and parallel pressures are expressed in terms of similar nonlinear FLR expansions. Lastly, we note that the treatment of higher-order gyrocenter-fluid moments (e.g., heat fluxes) is presently outside of the scope of a variational formulation (see Appendix A of Ref.~\cite{Brizard_2005} for additional comments).

\subsection{Energy conservation properties}

Third, the energy conservation properties of our nonlinear reduced fluid equations are guaranteed by the use of a variational principle. By using the Noether method \cite{Brizard_JPP}, the local energy conservation law
\begin{equation}
\pd{{\cal E}}{t} \;+\; \nabla\bdot{\bf S} \;=\; 0,
\label{eq:energy_local}
\end{equation}
is derived with explicit expressions for the energy density ${\cal E}$ and the energy-density flux ${\bf S}$. To verify energy conservation in our numerical simulations, we decompose the energy density ${\cal E} \equiv \sum_{i}\,{\cal E}_{i}$ in terms of the components ${\cal E}_{i}$ (with corresponding energy-flux decomposition ${\bf S} \equiv \sum_{i}\,{\bf S}_{i}$) and track the time evolution of each volume-integrated component $E_{i} \equiv \int_{V}\,{\cal E}_{i}\,d^{3}x$ in terms of the {\it energy-transfer} equations: 
\begin{equation}
\frac{dE_{i}}{dt} \;=\; \sum_{j}\,Q_{ij}, 
\label{eq:energy_i}
\end{equation}
where $Q_{ij} = -\,Q_{ji}$ denotes the (antisymmetric) energy transfer between components $i$ and $j$ (such that total energy conservation is guaranteed by $\sum_{i,j}\,Q_{ij} \equiv 0$) and boundary conditions are chosen such that the surface terms $\oint_{\partial V}\,{\bf S}_{i}\bdot\wh{{\sf n}}\,dA$ vanish in Eq.~(\ref{eq:energy_i}).

\subsection{Organization}

The remainder of the paper is organized as follows. In Sec.~\ref{sec:Brizard}, we review the derivation of the reduced fluid equations and the self-consistent Maxwell's equations of Ref.~\cite{Brizard_2005} by the Lagrangian variational method. The derivation differs from the one presented in Ref.~\cite{Brizard_2005} in our treatment of the perturbation magnetic field (\ref{eq:B_def}). In Sec.~\ref{sec:energy}, we derive the exact energy
conservation law (\ref{eq:energy_local}) by Noether's method. In Sec.~\ref{sec:reduced}, we rearrange the reduced parallel-force equation derived in Sec.~\ref{sec:Brizard} to display the nonlinear FLR effects explicitly based on Eq.~(\ref{eq:PhiA_rho}). In Sec.~\ref{sec:linearDispersionRelation}, we present a linear dispersion relation for a homogeneous two-fluid isotropic plasma. In Sec.~\ref{sec:simulationResults}, we present linear and nonlinear numerical simulations using our reduced-fluid equations in straight and magnetic dipole geometries and we summarize our work in Sec.~\ref{sec:summary}. 

In Appendix \ref{sec:push}, we present a summary of the general foundations of polarization and magnetization effects associated with dynamical reduction in plasma physics presented in Ref.~\cite{Brizard_2006} and, in Appendix \ref{sec:pol_disp}, we present a simple interpretation of the reduced-fluid displacement (\ref{eq:df_dipole}) as well as a simple derivation of the effective potentials (\ref{eq:PhiA_rho}).

\section{\label{sec:Brizard}Reduced-fluid equations}

The nonlinear finite-beta reduced-fluid equations derived by Brizard \cite{Brizard_2005} were obtained from a variational principle
\begin{equation}
\delta\;\int\;\L\;d^{3}x \;=\; 0,
\label{eq:pla}
\end{equation}
where the Lagrangian density (sum over species is implied unless otherwise noted) is
\begin{eqnarray}
\L & \equiv & \frac{1}{8\pi} \left( \left|{\bf E}_{\bot}\right|^{2} \;-\; |{\bf B}|^{2} \right) \;+\; qn\,\frac{{\bf u}}{c}\bdot\left( {\bf A}_{0} + 
A_{\|}\,\bhat_{0} \right) \label{eq:Lag_primitive} \\
 &  &+\;\frac{mn}{2}\;\left|u_{\|}\,\left(\bhat_{0} + \frac{{\bf B}_{\bot}}{B_{0}} \right) \;+\; {\bf u}_{E}\right|^{2}  \;-\; \left( qn\,\Phi \;+\; 
{\cal P} \right). \nonumber
\end{eqnarray}
Here, ${\cal P} = p_{\bot} + \frac{1}{2}\,p_{\|} \equiv \frac{1}{2}\,{\rm Tr}({\sf P})$ is the trace of the Chew-Goldberger-Low (CGL) pressure tensor 
\begin{equation}
{\sf P} \;=\; p_{\|}\,\bhat_{0}\bhat_{0} \;+\; p_{\bot}\,({\bf I} - \bhat_{0}\bhat_{0}). 
\label{eq:CGL_P}
\end{equation}
Note that the Lagrangian density (\ref{eq:Lag_primitive}) contains the linear $E\times B$ velocity ${\bf u}_{E}$ and the linear magnetic-flutter velocity \cite{flutter_1,flutter_2} $u_{\|}\,{\bf B}_{\bot}/B_{0}$ explicitly, while all other terms have standard interpretations. In addition, we note that the replacement $|{\bf E}|^{2} \rightarrow |{\bf E}_{\bot}|^{2}$ in the first term in Eq.~(\ref{eq:Lag_primitive}) removes the parallel displacement current  $(\partial_{t}E_{\|})$ in the parallel Amp\`{e}re equation and the total magnetic field is ${\bf B} = {\bf B}_{0} + {\bf B}_{\bot}$, where the background magnetic field ${\bf B}_{0} \equiv \nabla\btimes{\bf A}_{0}$ is assumed to be time-independent. 

Following Ref.~\cite{Brizard_2005}, the Lagrangian density (\ref{eq:Lag_primitive}) is also written as
\begin{eqnarray}
\L & = & \frac{1}{8\pi} \left( \left|{\bf E}_{\bot}\right|^{2} \;-\; |{\bf B}|^{2} \right) \;+\; \frac{qn}{c} \left( {\bf A}_{0}\bdot{\bf u} \;+\; 
u_{\|}\,A_{\|}^{*} \frac{}{}\right) \nonumber \\
 &  &+\;\frac{mn}{2}\;U_{\|}^{2}  \;-\; \left( qn\,\Phi^{*} \;+\; {\cal P} \right),
\label{eq:Lag_total}
\end{eqnarray}
where the effective fields are
\begin{equation}
\left. \begin{array}{rcl}
U_{\|} & \equiv & u_{\|} \left|\bhat_{0} + {\bf B}_{\bot}/B_{0}\right| \;\equiv u_{\|}\,b \\
 &  & \\
q\,\Phi^{*} & \equiv & q\,\Phi \;-\; m\,|{\bf u}_{E}|^{2}/2 \\
 &  & \\
(q/c)\,A_{\|}^{*} & \equiv & (q/c)\,A_{\|} \;-\; m\,V_{\|}
\end{array} \right\},
\label{eq:UPA_def}
\end{equation}
with the perturbed nonlinear $E\times B$ velocity
\begin{equation}
V_{\|} \;\equiv\; \frac{c\bhat_{0}}{B_{0}^{2}}\bdot\left({\bf E}_{\bot}\btimes{\bf B}_{\bot}\right) \;=\; -\;{\bf u}_{E}
\bdot\frac{{\bf B}_{\bot}}{B_{0}}.
\label{eq:ExB_nonlin}
\end{equation}
The parallel reduced-fluid velocity $U_{\|}$ represents motion along the perturbed magnetic field lines, the effective potential $\Phi^{*}$ includes the zero-Larmor-radius (ZLR) gyrokinetic electrostatic correction, and the effective potential $A_{\|}^{*}$ includes the parallel nonlinear $E\times B$ velocity 
\[ \frac{q}{c}\,A_{\|}^{*} \;\equiv\; \left(\bhat_{0} + \frac{{\bf B}_{\bot}}{B_{0}} \right) \bdot \left( \frac{q}{c}\,A_{\|}\,\bhat_{0} \;+\; m\,
{\bf u}_{E} \right). \]
We will see below that these definitions simplify the equations of motion obtained from the variational principle (\ref{eq:pla}).

It is useful to introduce the following nonlinear finite-Larmor-radius (FLR) identities involving the reduced-fluid displacement vector (\ref{eq:df_dipole}) and the definitions (\ref{eq:PhiA_rho}):
\begin{eqnarray}
q\,\Phi^{*} \;+\; \frac{m}{2}\,U_{\|}^{2} & = & q\,\Phi_{\rho} \;+\; \frac{m}{2}\,u_{\|}^{2} \;+\; K_{\rho}, \label{eq:phi_FLR} \\
\frac{q}{c}\,A_{\|}^{*} \;+\; mu_{\|}\,b^{2} & = & \frac{q}{c}\, A_{\|\rho} \;+\; mu_{\|}, 
\label{eq:A_FLR}
\end{eqnarray}
where the second-order Hamiltonian term
\begin{equation}
K_{\rho} \;\equiv\; \frac{m}{2}\,\Omega_{0}^{2}\,|\vb{\rho}_{\bot}|^{2}
\label{eq:K_rho}
\end{equation}
is interpreted as a low-frequency ponderomotive Hamiltonian term. Indeed, in nonlinear gyrokinetic theory \cite{Brizard_Hahm}, the second-order gyrocenter Hamiltonian term
\begin{eqnarray*} 
\frac{1}{2}\;\left\langle \left\{ S_{1},\; \{ S_{1},\; H_{0} \}\frac{}{} \right\} \right\rangle & = & 
\frac{\Omega_{0}^{2}}{2\,B_{0}}\, \pd{}{\mu}\left\langle \left( \pd{S_{1}}{\zeta}\right)^{2}\right\rangle \\
 & \equiv & \frac{m}{2}\;\Omega_{0}^{2} \left| \langle\vb{\rho}_{1}\rangle\right|^{2}
\end{eqnarray*}
yields, upon averaging with respect to the gyrocenter Vlasov distribution over gyrocenter momentum space, the relation
\begin{equation}
\frac{m}{2}\;\Omega_{0}^{2} \left| \langle\vb{\rho}_{1}\rangle\right|^{2} \;\rightarrow\; \frac{m}{2}\;\Omega_{0}^{2} \;|\vb{\rho}_{\bot}|^{2} \;\equiv\; K_{\rho},
\label{eq:H_pond}
\end{equation}
where we have neglected thermal effects. The form of this ponderomotive Hamiltonian is similar to the magnetic-moment Hamiltonian $\mu B \equiv 
\frac{1}{2}\,m\,\Omega_{0}^{2}\;\langle|\vb{\rho}_{0}|^{2}\rangle$ (i.e., the guiding-center ``ponderomotive" Hamiltonian). It is also similar to the high-frequency ponderomotive potential \cite{Skoric_Kono} $K_{{\rm w}} \equiv \frac{1}{2}\,m\,\omega^{\prime 2}\,\|\vb{\rho}_{{\rm w}}^{2}\|$, where 
$\|\cdots\|$ denotes eikonal-phase averaging and the high-frequency eikonal displacement $\wt{\vb{\rho}}_{{\rm w}} \equiv -\,(e/m\omega^{\prime 2})\,
(\wt{{\bf E}} + {\bf v}/c\btimes\wt{{\bf B}})$ is expressed in terms of the high-frequency wave electric and magnetic fields (with $\omega^{\prime} \equiv \omega - {\bf k}\bdot{\bf v}$).

\subsection{Reduced polarization and magnetization vectors}

From the Lagrangian density (\ref{eq:Lag_primitive}), we define the following reduced-fluid polarization and magnetization vectors
\begin{eqnarray}
{\bf P}_{\bot} & \equiv & \pd{\L}{{\bf E}_{\bot}} \;-\; \frac{{\bf E}_{\bot}}{4\pi} \;=\; \sum qn\,\vb{\rho}_{\bot},
\label{eq:Pbot_def} \\
{\bf M}_{\bot} & \equiv & \pd{\L}{{\bf B}_{\bot}} \;+\; \frac{{\bf B}_{\bot}}{4\pi} \;=\; \sum qn\,\left(\vb{\rho}_{\bot}\btimes \frac{u_{\|}}{c}
\bhat_{0} \right), \label{eq:Mbot_def}
\end{eqnarray}
which are expressed in terms of the reduced-fluid displacement vector (\ref{eq:df_dipole}). We note that the perpendicular magnetization vector 
(\ref{eq:Mbot_def}) is expressed in terms of the moving-dipole contribution only \cite{Jackson_1975}. Note also that the relation
\begin{equation}
  \label{eq:Krho_relation}
  \sum\,n\,K_{\rho} \;\equiv\; \frac{1}{2} \left( {\bf E}_{\bot}\bdot{\bf   
  P}_{\bot} \;+\; {\bf B}_{\bot}\bdot{\bf M}_{\bot} \right)
\end{equation}
exemplifies the K-$\chi$ Theorem \cite{Dewar,CK_1,CK_2}, which leads to the expression
\[ \frac{1}{8\pi} \left( \left|{\bf E}_{\bot}\right|^{2} \;-\; |{\bf B}|^{2} \right) \;+\; \sum\,n\,K_{\rho} \;\equiv\; \frac{1}{8\pi} \left( 
{\bf E}_{\bot}\bdot{\bf D} \;-\; {\bf B}\bdot{\bf H} \right), \]
where we have defined the macroscopic electromagnetic fields 
\begin{equation}
\left. \begin{array}{rcl}
{\bf D} & \equiv & {\bf E}_{\bot} \;+\; 4\pi\,{\bf P}_{\bot} \\
{\bf H} & \equiv & {\bf B} - 4\pi\,{\bf M}_{\bot}
\end{array} \right\}.
\label{eq:DH_def}
\end{equation}
The reduced Maxwell's equation are, thus, expressed as
\begin{equation} 
\nabla\bdot{\bf D} \;=\; 4\pi\;\sum\; q\; n,
\label{eq:D_Maxwell}
\end{equation}
and
\begin{equation}
\nabla\btimes{\bf H} \;-\; \frac{1}{c}\,\pd{{\bf D}}{t} \;=\; 4\pi\;\sum\;\frac{q}{c}\;n{\bf u},
\label{eq:H_Maxwell}
\end{equation}
where the right sides represent the total gyrocenter charge density and the total gyrocenter current density, respectively.

We may thus rewrite the Lagrangian (\ref{eq:Lag_total}) as
\begin{eqnarray}
\L & \equiv & \frac{1}{8\pi} \left( \left|{\bf E}_{\bot}\right|^{2} \;-\; |{\bf B}|^{2} \right) \;+\; n \left( \frac{m}{2}\;u_{\|}^{2} \;-\;
K_{\rho} \right) \nonumber \\
 &  &-\; {\cal P} \;+\; qn \left( {\bf A}_{0}\bdot\frac{{\bf u}}{c} + \frac{u_{\|}}{c}\,A_{\|\rho} - \Phi_{\rho} 
\right),
\label{eq:Lag_FLR}
\end{eqnarray}
where nonlinear FLR corrections (\ref{eq:PhiA_rho}) and the reduced-fluid ponderomotive (\ref{eq:H_pond}) are shown explicitly. 

In the absence of nonlinear FLR effects (i.e., $\vb{\rho}_{\bot} = 0$), the Lagrangian density (\ref{eq:Lag_FLR}) reverts back to the standard Lagrangian density for a guiding-center plasma fluid.

\subsection{Dynamical constraints}

The variational principle (\ref{eq:pla}) does not treat the fields $(n,{\bf u}, p_{\bot}, p_{\|}; {\bf E}, {\bf B})$ as independent variational fields. Instead, the Eulerian variations $(\delta n, \delta{\bf u}, \delta p_{\bot}, \delta p_{\|})$ are expressed in terms of the virtual fluid displacement $\vb{\xi}$ while the Eulerian variations $(\delta {\bf E}, \delta {\bf B})$ are expressed in terms of the potential variations $(\delta \Phi, \delta{\bf A})$ subject to constraint equations.

The constraint equations for the Eulerian fluid-moment variations $(\delta n, \delta{\bf u}, \delta p_{\bot}, 
\delta p_{\|})$ are the continuity equation for each reduced-fluid species
\begin{equation}
\pd{n}{t} \;=\; -\;\nabla\bdot(n\,{\bf u}),
\label{eq:continuity}
\end{equation}
and the CGL pressure equations
\begin{eqnarray}
\pd{p_{\bot}}{t} & = & -\,\nabla\bdot(p_{\bot}\,{\bf u}) \;-\; p_{\perp}\;({\bf I} - \bhat_{0}\bhat_{0}):\nabla{\bf u}, \label{eq:pperp} \\
\pd{p_{\|}}{t} & = & -\,\nabla\bdot(p_{\|}\,{\bf u}) \;-\; 2\,p_{\parallel}\;\bhat_{0}\bhat_{0}:\nabla{\bf u}, \label{eq:ppar}
\end{eqnarray}
where the higher-order heat-flux moments are omitted here (but were considered in Ref.~\cite{Brizard_2005}).

The Eulerian fluid-moment variations $\delta\eta^{a} = (\delta n, \delta p_{\bot}, \delta p_{\|})$ are defined in terms of the relations
\begin{equation} 
\delta\eta^{a} \;\equiv\; \lim_{\Delta t \rightarrow 0}\left( \pd{\eta^{a}}{t}\,\Delta t\right),
\label{eq:Delta_def}
\end{equation}
where the virtual fluid displacement $\vb{\xi}$ is defined as
\[ \vb{\xi} \;\equiv\; \lim_{\Delta t \rightarrow 0}\left( {\bf u}\,\Delta t\right). \]
According to Eqs.~(\ref{eq:continuity})-(\ref{eq:ppar}) and Eq.~(\ref{eq:Delta_def}), the Eulerian variations 
$\delta\eta^{a}$ are
\begin{equation}
\left. \begin{array}{rcl}
\delta n & = & -\;\nabla\bdot(n\;\vb{\xi}) \\
 &  & \\
\delta{\cal P} & = & -\;\nabla\bdot({\cal P}\;\vb{\xi}) \;-\; {\sf P}:\nabla\vb{\xi}
\end{array} \right\},
\label{eq:delta_np}
\end{equation}
and the Eulerian variation $\delta{\bf u} \equiv \Delta{\bf u} - \vb{\xi}\bdot\nabla{\bf u}$ of the fluid velocity is defined in terms of the Lagrangian variation
\[ \Delta{\bf u} \;\equiv\; \frac{d\vb{\xi}}{dt} \;=\; \pd{\vb{\xi}}{t} \;+\; {\bf u}\bdot\nabla\vb{\xi}. \]
Note that the Eulerian variations $\delta n$ and $\delta{\bf u}$ satisfy the constraint $\partial_{t}\delta n + \nabla\bdot(\delta n\,{\bf u} + n\,
\delta{\bf u}) = 0$, as expected.

The constraint equations for the Eulerian variations $(\delta {\bf E}, \delta {\bf B})$ are
\[ \nabla\btimes{\bf E} \;+\; \frac{1}{c}\,\pd{{\bf B}}{t} \;=\; 0 \;\;{\rm and}\;\; \nabla\bdot{\bf B} \;=\; 0. \]
The Eulerian variations for the electromagnetic fields ${\bf E}$ and ${\bf B}$:
\begin{equation}
\delta{\bf E} \;=\; -\;\nabla\delta\Phi \;-\; \frac{1}{c}\,\pd{\delta{\bf A}}{t} \;\;{\rm and}\;\;
\delta{\bf B} \;=\; \nabla\btimes\delta{\bf A}
\label{eq:delta_EB}
\end{equation}
are expressed in terms of the variations $\delta\Phi$ and $\delta{\bf A}$.

We express the variation of the Lagrangian (\ref{eq:Lag_total}) in terms of $(\vb{\xi}, \delta\Phi, \delta{\bf A})$ through the relations (\ref{eq:delta_np}) and (\ref{eq:delta_EB}) as
\begin{eqnarray}
\delta\L & = & -\;\vb{\xi}\bdot\left[ \pd{}{t}\left(\pd{\L}{{\bf u}}\right) + \nabla\bdot\left({\bf u}\;
\pd{\L}{{\bf u}} \right) + \nabla{\bf u}\bdot\pd{\L}{{\bf u}} \right. \nonumber \\
 &  &\left.+\; \nabla\bdot{\sf P} - \left( \eta^{a}\,\nabla\pd{\L}{\eta^{a}} \right) \right] + \delta\Phi \left( 
\pd{{\cal L}}{\Phi} + \nabla\bdot\pd{{\cal L}}{{\bf E}} \right) \nonumber \\
 &  &+\; \delta{\bf A}\bdot\left( \pd{{\cal L}}{{\bf A}} \;+\; \frac{1}{c}\;\pd{}{t}\;\pd{{\cal L}}{{\bf E}} \;+\; \nabla\btimes\pd{{\cal L}}{{\bf B}}
\right) \nonumber \\
 &  &+\; \pd{\delta\Lambda}{t} \;+\; \nabla\bdot\delta\vb{\Gamma},
\label{eq:deltaL_total}
\end{eqnarray}
where the space-time-divergence components
\begin{eqnarray}
\delta\Lambda & = & \vb{\xi}\bdot\pd{\L}{{\bf u}} \;-\; \frac{1}{c}\,\delta{\bf A}\bdot\pd{\L}{{\bf E}}, \label{eq:delta_Lambda} \\
\delta\vb{\Gamma} & = & {\bf u} \left( \vb{\xi}\bdot\pd{\L}{{\bf u}}\right) \;+\; \left( {\sf P} \;-\; \eta^{a}\,\pd{\L}{\eta^{a}}\;{\bf I} \right) 
\bdot\vb{\xi} \nonumber \\
 &  &-\; \left( \delta\Phi\;\pd{\L}{{\bf E}} \;+\; \delta{\bf A}\btimes\pd{\L}{{\bf B}} \right) \label{eq:delta_Gamma}
\end{eqnarray}
do not play a role in the least-action principle (\ref{eq:pla}) but, instead, play a crucial role in the derivation of exact conservation laws (in the next Section).

\subsection{Reduced-fluid equation of motion}

The stationarity of the action (\ref{eq:pla}) with respect to $\vb{\xi}$ yields the Euler-Poincar\'{e} equation for the reduced-fluid velocity ${\bf u}$:
\begin{eqnarray}
0 & = & \pd{}{t}\left(\pd{\L}{{\bf u}}\right) \;+\; \nabla\bdot\left({\bf u}\;
\pd{\L}{{\bf u}} \right) \;+\; \nabla{\bf u}\bdot\pd{\L}{{\bf u}} \nonumber \\
 &  &\mbox{}+\; \left( \nabla\bdot{\sf P} \;-\; n\;\nabla\pd{\L}{n} \right),
\label{eq:EP_u}
\end{eqnarray}
which becomes the reduced-fluid equation of motion
\begin{equation}
0 \;=\; mn\,\bhat_{0}\;\pd{u_{\|}^{*}}{t} \;-\; qn \left( {\bf E}^{*} \;+\; \frac{{\bf u}}{c}\btimes{\bf B}^{*} \right) 
\;+\; \nabla\bdot{\sf P}^{*},
\label{eq:gyro_velocity}
\end{equation}
where $u_{\|}^{*} \equiv u_{\|}\,b^{2}$ and $\nabla\bdot{\sf P}^{*} \equiv \nabla\bdot{\sf P} + \frac{1}{2}\;mn\,\nabla 
U_{\|}^{2}$, and we introduced the effective electric field
\begin{eqnarray}
{\bf E}^{*} & \equiv & -\;\nabla\Phi^{*} \;-\; \frac{\bhat_{0}}{c}\;\pd{A_{\|}^{*}}{t} \nonumber \\
 & = & {\bf E} \;+\; \frac{m}{q} \left( \frac{1}{2}\,\nabla|{\bf u}_{E}|^{2} \;+\; \pd{V_{\|}}{t}\;\bhat_{0} \right),
\label{eq:E_star}
\end{eqnarray}
and the effective magnetic field
\begin{eqnarray}
{\bf B}^{*} & \equiv & \nabla\btimes\left[\; {\bf A}_{0} \;+\; \left( A_{\|}^{*} \;+\; \frac{mc}{q}\,u_{\|}\,b^{2} 
\right) \bhat_{0}  \;\right] \nonumber \\
 & = & {\bf B}_{0} \;+\; \nabla \btimes\left[\; \left( A_{\|}^{*} \;+\; \frac{mc}{q}\,u_{\|}\,b^{2} 
\right) \bhat_{0} \;\right]. \label{eq:Bstar_def}
\end{eqnarray}
We clearly see that the definitions $(u_{\|}^{*}, \Phi^{*}, A_{\|}^{*})$ used to write Eq.~(\ref{eq:gyro_velocity}) in simple form actually hide all the nonlinear corrections.

The reduced-fluid equation (\ref{eq:gyro_velocity}) may be decomposed into two separate equations. First, the cross-product of 
Eq.~(\ref{eq:gyro_velocity}) with $\bhat_{0}$ yields the reduced-fluid velocity
\begin{equation}
{\bf u} \;\equiv\; u_{\|}\,\bstar \;+\; \frac{c\bhat_{0}}{qn\,B_{\|}^{*}}\btimes\left( \nabla\bdot{\sf P}^{*} \;+\;  qn\,
\nabla\Phi^{*} \right),
\label{eq:u_vector}
\end{equation}
where 
\begin{equation}
B_{\|}^{*} \;\equiv\; \bhat_{0}\bdot{\bf B}^{*} \;\;{\rm and}\;\; \bstar \;\equiv\; \frac{{\bf B}^{*}}{B_{\|}^{*}}.
\label{eq:bunit_star}
\end{equation}
Note that, under the assumption $\bhat_{0}\bdot\nabla\btimes\bhat_{0} = 0$, we find $B_{\|}^{*} = B_{0}$ and $\bstar = {\bf B}^{*}/B_{0}$. Next, the dot-product of Eq.~(\ref{eq:gyro_velocity}) with $\bstar$ yields the reduced-fluid parallel equation of motion
\begin{equation}
mn\;\pd{u_{\|}^{*}}{t} \;=\; \bstar\bdot\left( qn\,{\bf E}^{*} \;-\; \nabla\bdot{\sf P}^{*} \right).
\label{eq:u_parallel}
\end{equation}
Once again all nonlinear corrections are hidden in the definitions of the effective fields $(u_{\|}^{*}, \Phi^{*}, A_{\|}^{*})$, which may prevent us from arriving at a clear interpretation of these nonlinear terms. The simplicity of Eq.~(\ref{eq:u_parallel}), however, points toward some underlying principle behind the organization of the nonlinear terms. We postpone providing our new interpretation of Eq.~(\ref{eq:u_parallel}) in terms of nonlinear FLR corrections until Sec.~\ref{sec:reduced}.

\subsection{Reduced Maxwell's equations}

We now use our variational principle (\ref{eq:pla}), based on the variation (\ref{eq:deltaL_total}), to derive the reduced Maxwell's equations, which exhibit the important polarization and magnetization effects that provided the motivation for nonlinear gyrokinetic theory \cite{Brizard_Hahm}.

The stationarity of the action $\int \delta\L\,d^{3}x = 0$ with respect to $\delta\Phi$ yields the Euler-Poincar\'{e} equation
\begin{equation}
\pd{{\cal L}}{\Phi} \;+\; \nabla\bdot\pd{{\cal L}}{{\bf E}} \;=\; 0,
\label{eq:EP_phi}
\end{equation}
which becomes the reduced Poisson equation
\begin{equation}
\frac{\nabla\bdot{\bf E}_{\bot}}{4\pi} \;\equiv\; \sum\; q \left[\; n \;-\; \nabla\bdot(n\,\vb{\rho}_{\bot}) \;\right] \;\equiv\; \rho_{{\rm phys}}.
\label{eq:Poisson}
\end{equation}
Here, the physical charge density $\rho_{{\rm phys}}$ is expressed as the sum of the reduced charge density $(\sum\,en)$ and the polarization density $(-\,\nabla\bdot{\bf P}_{\bot})$.  

The stationarity of the action (\ref{eq:pla}) with respect to $\delta{\bf A}$ yields the Euler-Poincar\'{e} equation
\begin{equation}
\pd{{\cal L}}{{\bf A}} \;+\; \frac{1}{c}\;\pd{}{t}\;\pd{{\cal L}}{{\bf E}} \;+\; \nabla\btimes\pd{{\cal L}}{{\bf B}} \;=\; 0,
\label{eq:EP_A}
\end{equation}
which becomes the reduced Maxwell equation
\begin{eqnarray}
\nabla\btimes\frac{{\bf B}}{4\pi} & = & \sum\;\frac{q}{c}\,n{\bf u} \;+\; \frac{1}{c}\;\pd{}{t} \left( \frac{{\bf E}_{\bot}}{4\pi} \;+\; 
{\bf P}_{\bot} \right) \nonumber \\
 &  &+\; \nabla\btimes{\bf M}_{\bot} \;\equiv\; \frac{1}{c}\;{\bf J}_{{\rm phys}}.
\label{eq:Ampere}
\end{eqnarray}
Here, the physical charge current ${\bf J}_{{\rm phys}}$ is expressed as the sum of the reduced charge current $(\sum\,en{\bf u})$, the polarization current $(\partial{\bf P}_{\bot}/\partial t)$, and the magnetization current $(c\,\nabla\btimes{\bf M}_{\bot})$. The parallel component of Eq.~(\ref{eq:Ampere}) yields the reduced parallel-Amp\`{e}re equation
\begin{equation}
\bhat_{0}\bdot\nabla\btimes\left( \frac{{\bf B}_{\bot}}{4\pi} \;-\; {\bf M}_{\bot} \right) \;=\; \sum\; \frac{qn}{c}\;u_{\|}.
\label{eq:Ampere_par}
\end{equation}
Note that in a straight and uniform background magnetic field $(\nabla\btimes\bhat_{0} = 0)$, the reduced 
parallel-Amp\`{e}re equation (\ref{eq:Ampere_par}) becomes
\[ -\;\frac{\nabla_{\bot}^{2}A_{\|}}{4\pi} \;=\; \sum\,\frac{qn}{c}\; \left[\; n\,u_{\|} \;-\; \nabla\bdot(n\,u_{\|}\;\vb{\rho}_{\bot}) \;\right], \]
which appeared in Ref.~\cite{Brizard_2005}. The present work, however, makes use of the more general equation 
(\ref{eq:Ampere}).  In the special case of zero equilibrium current (such as for a dipole field), $\bhat_{0}\bdot\nabla\btimes{\bf B}/(4\pi) = -\;
\nabla\bdot[\nabla_\bot(A_\|/B_0)B_0^2]/(4\pi B_0)$.

\section{\label{sec:energy}Energy Conservation Laws}

One of the great advantages of using a variational principle to derive exact or reduced dynamical equations resides in the fact that these self-consistent equations are guaranteed to possess important conservation laws (e.g., energy-momentum or wave-action). This is especially important for reduced fluid equations that are derived by imposing an approximation scheme based on space-time-scale orderings on exact fluid equations.

When the three Euler-Poincar\'{e} equations (\ref{eq:EP_u}), (\ref{eq:EP_phi}), and (\ref{eq:EP_A}) are taken into account, the variation of the Lagrangian (\ref{eq:deltaL_total}) reduces to the Noether equation 
\begin{equation}
\delta\L \;=\; \pd{\delta\Lambda}{t} \;+\; \nabla\bdot\delta\vb{\Gamma}.
\label{eq:Noether}
\end{equation}
The Noether equation (\ref{eq:Noether}) may be used to derive conservation laws for the reduced fluid equations (\ref{eq:continuity})-({\ref{eq:ppar}), 
(\ref{eq:u_parallel}), (\ref{eq:Poisson}), and (\ref{eq:Ampere}), with the reduced-fluid velocity ${\bf u}$ given by Eq.~(\ref{eq:gyro_velocity}).

\subsection{Local energy conservation law}

We use the Noether equation (\ref{eq:Noether}) to derive a local energy conservation law associated with time-translation symmetry $(t \rightarrow t +
\delta t)$, where
\begin{equation}
\left. \begin{array}{rcl}
\vb{\xi} & = & -\;{\bf u}\;\delta t \\
\delta\Phi & = & -\;\delta t\;\partial_{t}\Phi \\
\delta{\bf A} & = & -\;\delta t\;\partial_{t}{\bf A} \;=\; c\delta t\;({\bf E} + \nabla\Phi) \\
\delta\L & = & -\;\delta t\;\partial_{t}\L
\end{array} \right\}.
\label{eq:delta_t}
\end{equation}
Upon rearranging terms and performing some gauge cancellations \cite{Brizard_2006,Brizard_JPP}, we obtain the local energy conservation law 
(\ref{eq:energy_local}), where the energy density is
\begin{equation}
{\cal E} \;\equiv\; {\bf u}\bdot\pd{\L}{{\bf u}} \;+\; \Phi\;\pd{\L}{\Phi} \;+\; {\bf E}\bdot\pd{\L}{{\bf E}} \;-\; \L,
\label{eq:E_density}
\end{equation}
and the energy-density flux is
\begin{eqnarray}
{\bf S} & \equiv & {\bf u} \left( {\bf u}\bdot\pd{\L}{{\bf u}} \right) \;+\; \left( {\sf P} \;-\; \eta^{a}\,\pd{\L}{\eta^{a}}\,{\bf I} \right)
\bdot{\bf u} \nonumber \\
 &  &-\; c \left( {\bf E}\btimes\pd{\L}{{\bf B}} \;+\; \Phi\;\pd{\L}{{\bf A}} \right).
\label{eq:E_flux}
\end{eqnarray}
By substituting derivatives of the reduced-fluid Lagrangian (\ref{eq:Lag_total}), Eqs.~(\ref{eq:E_density}) and (\ref{eq:E_flux}) become
\begin{eqnarray}
{\cal E} & = & -\;\frac{1}{8\pi} \left( |{\bf E}_{\bot}|^{2} \;-\; |{\bf B}|^{2} \right) \;+\; {\bf E}_{\bot}\bdot\frac{{\bf D}}{4\pi} \nonumber \\
 &  &+\; {\cal P} \;+\; \frac{mn}{2} \left( U_{\|}^{2} \;-\; |{\bf u}_{E}|^{2} \right) \nonumber \\
 & = & \frac{1}{8\pi} \left( |{\bf E}_{\bot}|^{2} \;+\; |{\bf B}|^{2} \right) \;+\; {\cal P} \nonumber \\
 &  &+\; \frac{mn}{2} \left| u_{\|}\,\left( \bhat_{0} + \frac{{\bf B}_{\bot}}{B_{0}} \right) \;+\; {\bf u}_{E}\right|^{2},
\label{eq:E_local}
\end{eqnarray}
and
\begin{eqnarray}
{\bf S} & = & \frac{c}{4\pi}\;{\bf E}_{\bot}\btimes{\bf H} \;+\; \left( {\sf P} \;+\; {\cal P}\,{\bf I} \right)\bdot{\bf u} \nonumber \\
 &  &+\; {\bf u} \left[\; \frac{mn}{2}\;\left( U_{\|}^{2} \;-\; |{\bf u}_{E}|^{2} \right) \;\right],
\label{eq:S_local}
\end{eqnarray}
which are identical to those presented by Brizard \cite{Brizard_2005} if we take into account that the perturbed magnetic field 
${\bf B}_{\bot}$ is now divergenceless.

\subsection{\label{subsec:global}Global energy conservation law}

By combining the CGL pressure equations (\ref{eq:pperp}) and (\ref{eq:ppar}), we obtain the evolution equation for the internal (pressure) energy density
\begin{equation}
\pd{{\cal P}}{t} \;=\; -\nabla\bdot\left( {\bf u}\,{\cal P} + {\sf P}\bdot{\bf u} \right) \;+\; {\bf u}\bdot(\nabla\bdot{\sf P}),
\label{eq:calP_dot}
\end{equation}
where the energy-flux term appears in Eq.~(\ref{eq:S_local}). By combining the continuity equation (\ref{eq:continuity}) with the reduced parallel-acceleration equation (\ref{eq:u_parallel}), we obtain the evolution equation for the parallel kinetic energy density (which includes motion along perturbed magnetic field lines) 
\begin{eqnarray}
\pd{}{t} \left( \frac{mn}{2}\,U_{\|}^{2} \right) & = & -\;\nabla\bdot\left[\; {\bf u} \left( \frac{mn}{2}\,U_{\|}^{2} \right) \;\right] \;-\;
{\bf u}\bdot(\nabla\bdot{\sf P}) \nonumber \\
 &  &+\; n{\bf u}\bdot \left( q\,{\bf E}^{*} \;-\; m\,U_{\|}\bhat_{0}\;\pd{b}{t} \right),
\label{eq:Upar_dot}
\end{eqnarray}
where the energy-flux term appears in Eq.~(\ref{eq:S_local}). Lastly, by dotting the reduced Maxwell equation (\ref{eq:Ampere}) with ${\bf E}_{\bot}$, we obtain the evolution equation for the electromagnetic energy density
\begin{eqnarray}
 &  &\pd{}{t} \left[\; {\bf E}_{\bot}\bdot\frac{{\bf D}}{4\pi} \;-\; \frac{1}{8\pi} \left( |{\bf E}_{\bot}|^{2} \;-\; |{\bf B}|^{2} \right) \;-\;
\frac{mn}{2}\,|{\bf u}_{E}|^{2} \;\right] \nonumber \\
 & = & -\;\nabla\bdot\left[\; \frac{c}{4\pi}\,{\bf E}_{\bot}\btimes{\bf H} \;-\; {\bf u} \left( \frac{mn}{2}\,|{\bf u}_{E}|^{2} \right) \;\right]
\nonumber \\
 &  &-\; n{\bf u}\bdot \left( q\,{\bf E}^{*} \;-\; m\,U_{\|}\bhat_{0}\;\pd{b}{t} \right).
\label{eq:Maxwell_dot}
\end{eqnarray}
where the energy-flux term appears in Eq.~(\ref{eq:S_local}). 

By adding the evolution equations (\ref{eq:calP_dot})-(\ref{eq:Maxwell_dot}), all transfer terms (i.e., terms that are not exact divergences) cancel each other exactly and we recover the local energy conservation law (\ref{eq:energy_local}), where the energy density ${\cal E}$ and energy-density flux ${\bf S}$ are given by Eqs.~(\ref{eq:E_local}) and (\ref{eq:S_local}), respectively. We note that, if we label the evolution equations 
(\ref{eq:calP_dot})-(\ref{eq:Maxwell_dot}) as $\partial{\cal E}_{i}/\partial t + \nabla\bdot{\bf S}_{i} = \sum_{j \neq i}\;q_{ij}$ with $i = 1$ (internal energy), $i = 2$ (parallel kinetic energy), and $i = 3$ (electromagnetic energy), then the antisymmetric energy-transfer density matrix 
$q_{ij} = -\,q_{ji}$ has the nonzero components
\begin{eqnarray}
q_{12} & = & {\bf u}\bdot(\nabla\bdot{\sf P}), \label{eq:q_12} \\
q_{23} & = & n{\bf u}\bdot \left( q\,{\bf E}^{*} \;-\; m\,U_{\|}\bhat_{0}\;\pd{b}{t} \right). \label{eq:q_23}
\end{eqnarray}
The energy-transfer equations (\ref{eq:energy_i}) therefore become
\begin{equation}
\frac{d}{dt} \left( \begin{array}{c}
E_{1} \\
E_{2} \\
E_{3}
\end{array} \right) \;\equiv\; \left( \begin{array}{c}
Q_{12} \\
-\,Q_{12} \;+\; Q_{23} \\
-\,Q_{23}
\end{array} \right),
\label{eq:energy_123}
\end{equation}
where $Q_{ij} \equiv \int_{V}\;q_{ij}\,d^{3}x = -\,Q_{ji}$ denotes a component of the volume-integrated energy-transfer matrix. Equation 
(\ref{eq:energy_123}) shows the importance of the effective parallel kinetic energy $(E_{2})$ in the transfer processes with the internal (pressure) energy $(E_{1})$ and the electromagnetic energy $(E_{3})$.

\section{\label{sec:reduced}NFLR Effects in Reduced-fluid Parallel Dynamics}

The reduced-fluid parallel equation of motion (\ref{eq:u_parallel}) gives the time evolution of the effective parallel velocity field $u_{\|}^{*} \equiv u_{\|}\,b^{2} = u_{\|}\,(1 + |{\bf B}_{\bot}|^{2}/B_{0}^{2})$ in terms of the effective fields (\ref{eq:UPA_def}). Equation (\ref{eq:u_parallel}) is written in a form where the convective part ${\bf u}\bdot\nabla u_{\|}$ is hiding on the right side. This equation may be written in a more standard form that brings out the NFLR corrections (\ref{eq:PhiA_rho}) explicitly. In order to facilitate our interpretation of this equation, it is preferable to write it in a way that explicitly displays the total time derivative, $du_{\|}/dt = \partial_{t}u_{\|} + {\bf u}\bdot\nabla u_{\|}$, of the parallel reduced-fluid velocity $u_{\|}$. 

First, using the nonlinear FLR identities (\ref{eq:phi_FLR}) and (\ref{eq:A_FLR}), Eq.~(\ref{eq:u_parallel}) becomes
\begin{equation}
n\;\pd{}{t} \left( m\,u_{\|} \;+\; \frac{q}{c}\;A_{\|\rho} \right) \;=\; \bstar\bdot\left({\bf F}_{\rho} \;-\; mn\;u_{\|}\nabla u_{\|} \right),
\label{eq:u_rho}
\end{equation}
where we introduced the NFLR-corrected force density
\begin{eqnarray}
{\bf F}_{\rho} & \equiv & -\;\nabla\bdot{\sf P} \;-\; n \nabla\left( q\;\Phi_{\rho} \;+\; K_{\rho} \right) \nonumber \\
 & = & -\;\nabla\bdot{\sf P}_{\rho} \;-\; qn \nabla\Phi_{\rho}.
\label{eq:F_rho}
\end{eqnarray}
The effective CGL-pressure force density
\begin{equation}
\nabla\bdot{\sf P}_{\rho} \;\equiv\; \nabla\bdot{\sf P} \;+\; n\;\nabla K_{\rho}
\label{eq:CGL_rho}
\end{equation}
now contains the low-frequency ponderomotive force density $n\,\nabla K_{\rho}$. This ponderomotive correction appears in complete analogy with the high-frequency ponderomotive force density \cite{Landau} that appears in reduced fluid models \cite{SK,SKH}. 

Next, we write the magnetic vector (\ref{eq:bunit_star}) as
\begin{equation}
\bstar \;\equiv\; {\sf b}^{*}_{\rho} \;+\; \nabla u_{\|}\btimes\frac{\bhat_{0}}{\Omega_{0}},
\label{eq:b_rho}
\end{equation}
where
\begin{equation}
{\sf b}^{*}_{\rho} \;\equiv\; \bhat_{0} \;+\; \frac{u_{\|}}{\Omega_{0}}\;\nabla\btimes\bhat_{0} \;+\; \frac{{\bf B}_{\bot \rho}}{B_{0}}
\label{eq:brho_star}
\end{equation}
includes the standard guiding-center curvature term $(u_{\|}/\Omega_{0})\,\nabla\btimes\bhat_{0}$ and the NFLR-corrected perturbed magnetic field
\begin{equation}
{\bf B}_{\bot \rho} \;\equiv\; \nabla\btimes\left( A_{\|\rho}\;\bhat_{0}\right),
\label{eq:Bbot_rho}
\end{equation}
so that the reduced-fluid velocity (\ref{eq:u_vector}) may be written as
\begin{equation}
{\bf u} \;=\; u_{\|}\,{\sf b}^{*}_{\rho} \;+\; \frac{{\bf F}_{\rho}}{mn}\btimes\frac{\bhat_{0}}{\Omega_{0}}.
\label{eq:uvec_rho}
\end{equation}

After rearranging terms in Eq.~(\ref{eq:u_parallel}) and using the identity
\begin{eqnarray*} 
-mn\,u_{\|}\bstar\bdot\nabla u_{\|} & = & -mn\,u_{\|}{\sf b}^{*}_{\rho}\bdot\nabla u_{\|} \\
 & = & -mn\,{\bf u}\bdot\nabla u_{\|} \;-\; \frac{\bhat_{0}}{\Omega_{0}}\btimes\nabla u_{\|}\bdot{\bf F}_{\rho} \\
 & \equiv & -mn\,{\bf u}\bdot\nabla u_{\|} \;+\; ({\sf b}_{\rho}^{*} - \bstar)\bdot{\bf F}_{\rho},
\end{eqnarray*}
we finally obtain
\begin{eqnarray}
mn\;\frac{du_{\|}}{dt} & = & {\sf b}_{\rho}^{*}\bdot\left( {\bf F}_{\rho} \;-\; \frac{qn}{c}\,\bhat_{0}\;\pd{A_{\|\rho}}{t} \right) \nonumber \\ 
 & \equiv & {\sf b}_{\rho}^{*}\bdot\left( qn\;{\bf E}_{\rho} \;-\; \nabla\bdot{\sf P}_{\rho} \right),
\label{eq:uparallel_rho}
\end{eqnarray}
where the NFLR-corrected electric field is
\begin{equation}
{\bf E}_{\rho} \;\equiv\; -\;\nabla\Phi_{\rho} \;-\; \frac{\bhat_{0}}{c}\;\pd{A_{\|\rho}}{t}.
\label{eq:E_rho}
\end{equation}
It is immediately clear that in the absence of nonlinear FLR and ponderomotive effects (i.e., $\vb{\rho}_{\bot} \equiv 0$) and omitting background magnetic curvature, the reduced-fluid parallel equation of motion (\ref{eq:uparallel_rho}) reverts back to the standard parallel equation
\[ mn\,\frac{du_{\|}}{dt} \;=\; \left( \bhat_{0} \;+\; \frac{{\bf B}_{\bot}}{B_{0}} \right)\bdot \left( qn\;{\bf E} 
\;-\; \nabla\bdot{\sf P} \right), \]
where the total time derivative $d/dt = \partial/\partial t + {\bf u}\bdot\nabla$ is expressed in terms of the guiding-center fluid velocity
\[ {\bf u} \;=\; u_{\|} \left( \bhat_{0} \;+\; \frac{{\bf B}_{\bot}}{B_{0}} \right) \;+\; \frac{c\bhat_{0}}{qnB_{0}}
\btimes\left( qn\;\nabla\Phi \;+\; \nabla\bdot{\sf P} \right), \]
which includes the standard $E\times B$ and magnetic-flutter convective nonlinearities. With the nonlinear FLR and ponderomotive effects retained, 
Eq.~(\ref{eq:uparallel_rho}) expresses the parallel momentum equation in the displaced (by $\vb{\rho}_{\bot}$) frame of the gyrocenters. 

\section{\label{sec:linearDispersionRelation} Linear Dispersion Relation}

For a two-component plasma fluid in a homogeneous magnetic field (for which the equations of this paper are identical to those of Brizard \cite{Brizard_2005}), the linear dispersion relation of our equations is expressed (in terms of the symbols defined in Table \ref{tab:sim_par}) as
\begin{eqnarray}
   \left( V - \frac{1}{1 + \ec^2 + \epsm} \right)
 \left[ V \left( 1 + \epsm \right) - \betatotp \right] \nonumber \\
   +  \left[ \epsm V^2 - V \left( \betaep + \epsm \betaip \right) + \betaep \betaip \right] \kappa = 0,
   \label{E:lindisp}
\end{eqnarray}
where $V_A \equiv B_0 / \sqrt{ 4 \pi m_i n_i }$ is the \Alfven\ speed, $\omega_{pi} \equiv \sqrt{ 4 \pi n_i q_i^2 / m_i}$ is the ion plasma frequency, 
$\beta_s = 8 \pi p_s / B_0^2$ is the plasma beta for species $s$ ($= i$ or $e$ for ions or electrons), and $\gamma_\parallel \;(= 3)$ is the ratio of specific heats for the parallel (1D) motion.  

\begin{widetext}

\begin{table}
\caption{\label{tab:sim_par}Parameter definitions and values for the linear dispersion relation (Fig.~\ref{F:LinearDispersion}), the linear simulation in straight geometry (Fig.~\ref{F:LinEnCon}), and the linear and nonlinear simulations (Figs.~\ref{F:LinSimResults} \& \ref{F:NonlinEnCon}) in dipole geometry.} 
\begin{ruledtabular} 
\begin{tabular}{ccccc}
Symbol                 & Definition                                 & Figure 1                & Figure 2 & Figures 4 \& 5 \\ \hline
$V$                    & $(\omega/(\kpl V_A))^2$                      & --                      & --       & -- \\ 
$\kappa$                    & $(k_{\bot}c/\omega_{pi})^{2}$              & --                      & $1$      & $0.001$ \\ 
$\epsilon_{c}$         & $V_{A}/c$                                  & $0.01$                  & $0.05$   & $0.02$ \\ 
$\epsilon_{m}$         & $m_{e}/m_{i}$                              & $0.001$                 & $0.01$   & $0.001$ \\ 
$\beta_{s}^{\prime}$   & $3\,\beta_{s}/2$                           & $0.05$                  & $0.15$   & $0.045$ \\ 
$\beta_{{\rm tot}}^{\prime}$ & $\beta_{i}^{\prime} + \beta_{e}^{\prime}$  & $0.1$                   & $0.3$    & $0.09$ \\ 
\end{tabular}
\end{ruledtabular}
\end{table}

\end{widetext}

The first line of Eq.~(\ref{E:lindisp}) (neglecting the terms proportional to $\kappa$, and with $\em$ and $\ec$ small) yields \Alfven\ wave ($V \sim 1$) and sound wave ($V \sim \betatotp$) solutions.  These basic waves are modified by finite $\kappa$, $\em$, and $\ec$. Further details concerning the interpretation of Eq.~(\ref{E:lindisp}) are given in Ref.~\citep{DentonEA07,footnote_2}. 

\begin{figure}
\includegraphics[width=18pc]{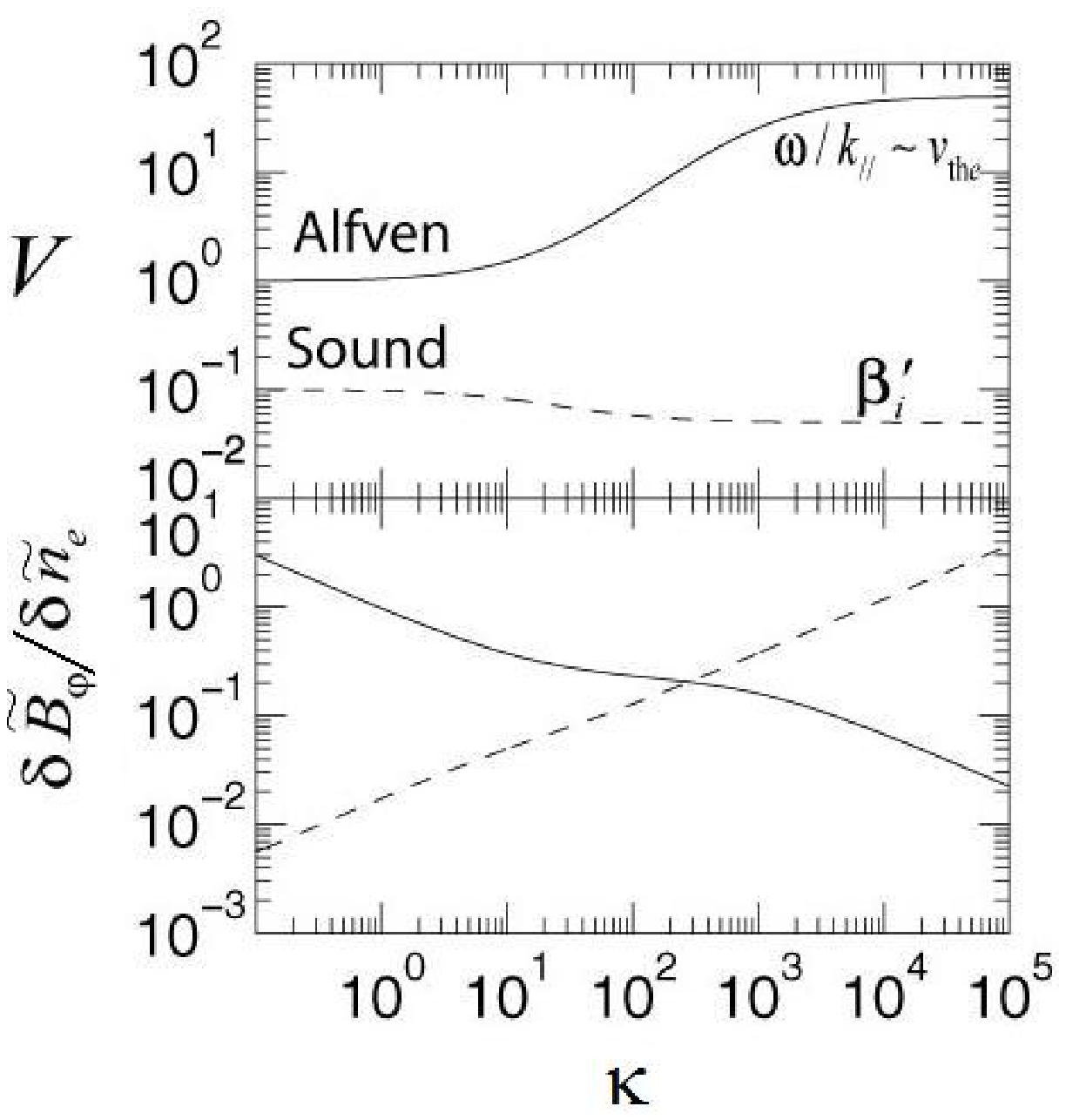} 
\caption{\label{F:LinearDispersion} Plotted versus $\kappa$ (normalized $k_\perp^2$ defined in Table \ref{tab:sim_par}), (a) the normalized squared phase speed $V$ and (b) the ratio of relative magnetic fluctuations to relative density fluctuations 
  $\delta\wt{B}_y/\delta\wt{n}_e \equiv (\delta B_y / B_0)/(\delta n_e / n_0)$ for the linear 
  dispersion relation Eq.~(\ref{E:lindisp}) for a homogeneous two-fluid isotropic plasma   
  with parameters defined in Table \ref{tab:sim_par}.  
}
\end{figure}

Figure~\ref{F:LinearDispersion} shows the normalized squared phase speed (defined in Table \ref{tab:sim_par} by $V$) and the ratio of relative magnetic fluctuations to relative density fluctuations $\delta\wt{B}_y/\delta \wt{n}_e = (\delta B_y / B_0)/(\delta n_e / n_0)$ versus the normalized squared perpendicular wavevector $\kappa$ (defined in Table \ref{tab:sim_par}, with $k_{\bot} = k_{x}$) using the linear dispersion relation Eq.~(\ref{E:lindisp}).  Here, $\delta n_e = n_e - n_0$ is the perturbed gyrocenter density rather than the perturbed physical density $(\delta n_{{\rm phys}})_{e} = n_e - \nabla\bdot(n_e\vb{\rho}_{\bot e}) - n_0$ [Eq.~(\ref{eq:red_phys})].  However, for the electrons, 
$\nabla\bdot(n_e\vb{\rho}_{\bot e})$ is small (because $\vb{\rho}_{\bot e} \propto m_e$) so $\delta n_e$ is approximately equal to the perturbed physical density.  (On the other hand, $n_i$ is not a good approximation for the physical ion density unless the density perturbations are dominated by fluctuations caused by parallel motion, as for low beta sound waves.)  Note in Fig.~\ref{F:LinearDispersion} that at low $\kappa$, there are 
\Alfven\ (solid curve) and sound wave (dashed curve) solutions.  The Alfven solution has $V = 1$ ($\omega / k_\parallel = V_A$) and small density fluctuations relative to magnetic fluctuations.  The sound wave solution has $V = \betatotp$ [$(\omega / k_\parallel)^2$ equal to the squared sound speed $\gamma_\parallel (T_i + T_e)/m_i$] and small magnetic fluctuations compared to density fluctuations.  Unlike the MHD equations, our equations have coupled parallel fluid motion and parallel current [Eq.~(\ref{eq:Ampere})], so there are coupled magnetic and density perturbations for both waves.  

At $\kappa \beta'_e \sim 1$, the $V \beta'_e \kappa$ term in Eq.~(\ref{E:lindisp}) starts to play a role.  The phase speed of the \Alfven\ wave starts to increase in a way characteristic of the kinetic \Alfven\ wave \citep{RogersEA01} and the character of the fluctuations (magnetic or 
acoustic) reverses for the two solutions.  At very large $\kappa$ [for which our equations are probably not accurate (unless $\beta_i \ll \beta_e$) because they do not include standard FLR corrections, e.g., $(k_{\bot}\rho_{i})^{2} = \frac{1}{2}\,\kappa\,\beta_{i} \ll 1$], the solution on the \Alfven\ wave branch (solid curve) becomes an electron sound wave with $\omega / k_\parallel = v_{{\rm th}_e}$, the electron thermal speed.  On the other hand, the solution on the sound wave branch has phase velocity equal to the sound speed based only on the ion temperature.

\section{\label{sec:simulationResults}Simulation Results}

\begin{table}
\caption{\label{tab:equations}Reduced fluid equations used in the 2-D reduced MHD simulation code.} 
\begin{ruledtabular} 
\begin{tabular}{ll}
Continuity for each species             & Eq.~(\ref{eq:continuity}) \\ 
Parallel momentum for each species      & Eq.~(\ref{eq:u_parallel}) \\ 
Perpendicular pressure for each species & Eq.~(\ref{eq:pperp}) \\ 
Parallel pressure for each species      & Eq.~(\ref{eq:ppar}) \\ 
Reduced-fluid velocity for each species & Eq.~(\ref{eq:u_rho}) \\ 
Reduced fluid displacement              & Eq.~(\ref{eq:df_dipole}) \\ 
Effective potentials                    & Eq.~(\ref{eq:UPA_def}) \\ 
Perturbed electric field                & Eq.~(\ref{eq:E_def}) \\ 
Perturbed magnetic field                & Eq.~(\ref{eq:B_def}) \\ 
Reduced Poisson equation                & Eq.~(\ref{eq:Poisson}) \\ 
Reduced Amp\`{e}re equation             & Eq.~(\ref{eq:Ampere}) \\ 
\end{tabular}
\end{ruledtabular}
\end{table}

We have implemented a two-dimensional reduced magnetohydrodynamic (MHD) finite-difference simulation using the equations in this paper (see Table 
\ref{tab:equations}). One of the two dimensions of the simulation is the direction of the background magnetic field. The code uses generalized orthogonal coordinates \citep{Arfken70}, and is second-order accurate with respect to time and fourth-order accurate with respect to space. For this paper, we use an insulator boundary at the ends of the simulation encountered by moving along the background magnetic field, and a hard-wall perfect conductor boundary at the ends of the simulation encountered by moving within the simulation plane perpendicular to the background magnetic field \citep{DentonEA08}.  These boundary conditions are energy conserving in the sense that there is no flux of energy out the boundaries. There is one modification of the equations that we made in the simulation code.  We set $b = 1$ [defined in Eq.~(\ref{eq:UPA_def})] and dropped the magnetic term in $\rho_\bot$ 
[Eq.~(\ref{eq:df_dipole})] only within the magnetization current (\ref{eq:Mbot_def}) that appears in the reduced parallel-Ampere equation 
(\ref{eq:Ampere_par}).  Making both of these changes together still maintains energy conservation.  The assumption is that the change in the total magnetic field amplitude caused by the (perpendicular) perturbation of the magnetic field is small, an assumption that is well satisfied for a low 
beta plasma such as occurs in the dipole magnetosphere at low altitudes.

The main purpose of the simulations is to demonstrate good energy conservation, a major advantage of our Lagrangian formulation. Note that the standard convective $E\times B$ and magnetic-flutter nonlinearities vanish in two dimensions (since ${\bf u}_{{\rm E}}\bdot\nabla = 0 = {\bf B}_{\bot}\bdot
\nabla$), which leaves only the parallel convective term $u_{\|}\bhat_{0}\bdot\nabla \neq 0$. The nonlinear FLR corrections, on the other hand, involve the differential operator
\[ \vb{\rho}_{\bot}\bdot\nabla \;=\; \frac{\bhat_{0}}{\Omega_{0}}\btimes\left( {\bf u}_{{\rm E}} \;+\; u_{\|}\;\frac{{\bf B}_{\bot}}{B_{0}}\right)\bdot
\nabla \;\neq\; 0, \]
which does not vanish in two-dimensional geometry. The use of a two-dimensional simulation geometry therefore enables us to focus our attention on the nonlinear FLR effects considered in the present work.

The energy density ${\cal E}$ is given in Eq.~(\ref{eq:E_local}).  We write 
\begin{equation}
  {\cal E} = {\cal E}_B + {\cal E}_E + {\cal E}_{K\parallel} + 
   {\cal E}_{K\perp} + {\cal E}_{P\parallel} + {\cal E}_{P\perp},
  \label{E:endef}
\end{equation} 
with
\begin{subequations}
 \label{E:endef_terms}
  \begin{eqnarray}
    {\cal E}_B & \equiv & \frac{1}{8\pi}|{\bf B}_{\bot}|^2,  \label{E:endef_B} \\
    {\cal E}_E & \equiv & \frac{1}{8\pi}|{\bf E}_{\bot}|^2,  \label{E:endef_E} \\
    {\cal E}_{K\parallel} & \equiv & \frac{1}{2}\;\sum_s m_s n_s u_{\parallel s}^2, \label{E:endef_Kpl} \\
   {\cal E}_{K\perp} & \equiv & \frac{1}{2}\;\sum_s m_s n_s \left|{\bf u}_{E} +  u_{\parallel s} \frac{{\bf B}_{\bot}}{B_0} \right|^2, \label{E:endef_Kpr} \\
   {\cal E}_{P\parallel} & \equiv & \frac{1}{2}\;\left.\left. \sum_s  \right( p_{\parallel s} \;-\; p_{\parallel s0} \right), \label{E:endef_Ppl} \\
   {\cal E}_{P\perp} & \equiv & \left.\left. \sum_s \right( p_{\perp s} \;-\; p_{\perp s0} \right). \label{E:endef_Ppr}
\end{eqnarray}
\end{subequations}
Note that the perturbation fields ${\bf E}_{\bot}$ and ${\bf B}_{\bot}$ are physical fields, the CGL-pressure components $(p_{\bot s}, p_{\| s})$ are physical fields in the standard zero-Larmor-radius limit [$(p_{\bot s0}, p_{\| s0})$ denote initial (equilibrium) values], and the fields $(n_{s}, u_{\| s})$ are gyrocenter-fluid fields related to their corresponding physical fields by Eq.~(\ref{eq:red_phys}). This division of terms is somewhat different than that of Sec.~\ref{subsec:global}, where the energy density is divided into terms that best demonstrate the pathways of energy flow.  Here, we chose to express the energy density terms in such a way that each term is positive definite [except for Eqs.~(\ref{E:endef_Ppl})-(\ref{E:endef_Ppr})].  For the results shown in the following simulation plots, the energies are volume averaged over the system volume [see Eq.~(\ref{eq:energy_i})].  Using the finite difference method, this means that we add up the products of the energy density and grid cell volume at each grid point. The resulting quantities are energies rather than energy density, but for the rest of this section, we use the same variable names defined in Eqs.~(\ref{E:endef_terms}) (labels for the energy density terms) for the energy terms.

\subsection{Linear simulation in straight geometry}

\begin{figure}
\includegraphics[width=15pc]{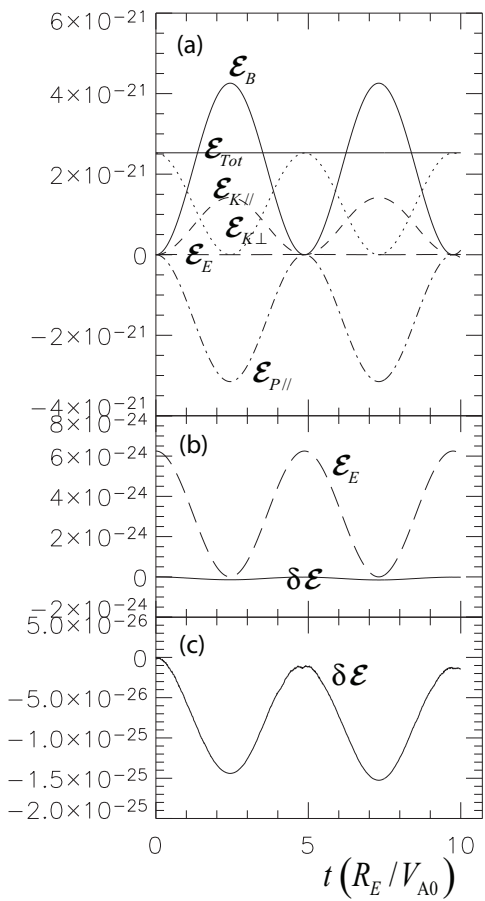} 
\caption{\label{F:LinEnCon} Energy terms (as described in the text) versus time for   
  a linear ($\delta v_y / V_A = 10^{-10}$) simulation in straight 
  geometry with parameters given in Table \ref{tab:sim_par}.  We plot in (a), all the 
  energy terms except the change in the total energy $\delta {\cal E}$, 
  in (b), $\delta {\cal E}$ and the electric field (displacement 
  current) energy ${\cal E}_E$, and in (c), just $\delta {\cal E}$.
}
\end{figure}

Figure~\ref{F:LinEnCon} shows energy terms [based on energy densities defined in Eqs.~(\ref{E:endef_terms})] and the change in the total energy from the beginning of the run, $\delta {\cal E}$, versus time $t$ for a linear ($\delta v_y / V_A = 10^{-10}$) simulation in straight geometry.  The parameters are given in Table \ref{tab:sim_par}.  The energy terms are integrated over space and normalized so that the energy of the constant equilibrium magnetic field $B_0$ would be unity; time is normalized to an arbitrary normalization length (written in the figure caption as $R_E$) divided by the \Alfven\ speed.  The code was initialized with a sinusoidal (in both simulation directions) wave perturbation of the out-of-plane velocity ${\bf u}_{E}$ consistent with a wave on the \Alfven\ branch in Fig.~\ref{F:LinearDispersion} (solid curve).  For these parameters, the total time of the simulation run was equal to the wave period of the wave determined from Eq.~(\ref{E:lindisp}) on the \Alfven\ branch [plus sign of quadratic equation for $V$ in 
Eq.~(\ref{E:lindisp})].  

Two major results can be seen from Fig.~\ref{F:LinEnCon}.  First, the resulting oscillations are consistent with the wave period from 
Eq.~(\ref{E:lindisp}) since the run has two complete oscillations of the (quadratic) energy terms.  Secondly, the change in the total energy 
(labeled $\delta {\cal E}$) is much smaller than the change in any individual energy term as can be seen by comparing the size of the terms in 
Fig.~\ref{F:LinEnCon}a, b, and c.  Whereas rough constancy of the total energy (Fig.~\ref{F:LinEnCon}a) does not well demonstrate energy conservation, the fact that the change in the total energy is smaller than the change in energy of the individual terms (Fig.~\ref{F:LinEnCon}a and~b) does usually indicate good energy conservation.  It shows that the error in the energy is smaller than the energy associated with the dynamics of the particular term.  Fig.~\ref{F:LinEnCon}b shows that the change in the total energy is even less than the energy of the electric field associated with the displacement current (usually neglected).  In Sec.~\ref{S:nonlinSim}, we present a more detailed convergence study demonstrating the quality of the energy conservation.

\subsection{Linear simulation in dipole geometry}

\begin{figure}
\includegraphics[width=15pc]{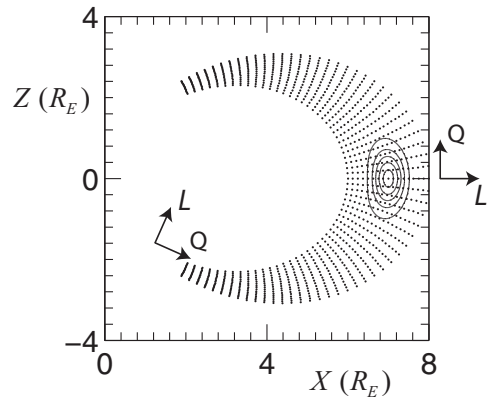} 
\caption{\label{F:DipoleGrid} Grid points (dots) of the 
  dipole simulation plotted in real (Cartesian) coordinates $X$ and 
  $Z$ (direction of dipole axis). [Only every fourth point is plotted in 
  the $Q$ (parallel) direction and every sixteenth point in the $L$ 
  (across field) direction.]  
  The curvilinear coordinate 
  directions are also shown, with the parallel coordinate $Q$ varying along the 
  magnetic field and the coordinate $L$ varying across the magnetic field.  Also 
  shown are contours for the initial perturbation in $\Phi$ centered at 
  ($X$,$Z$) = (7,0).
}
\end{figure}

Next, we show linear results in dipole geometry.  The simulation grid and coordinate system are described in Fig.~\ref{F:DipoleGrid}.  We use 
a $256 \times 256$ grid in the coordinates $Q$ and $L$ (Fig.~\ref{F:DipoleGrid}).  The coordinate $L$ is the $L$ shell used in magnetospheric physics, and is equal to the radial distance at the magnetic equator ($Z = 0$ in Fig.~\ref{F:DipoleGrid}) in units of the Earth radius $R_E$.  The system is perturbed with a peaked distribution of $\Phi$ at $L = 7$ at the magnetic equator (contours in Fig.~\ref{F:DipoleGrid}).  The parameters of this 
two-fluid plasma simulation are given in Table \ref{tab:sim_par} (where $\kappa$ is defined at the magnetic equator based on the scale length of the initial value for $\Phi$).  

\begin{figure*}    
\includegraphics[width=30pc]{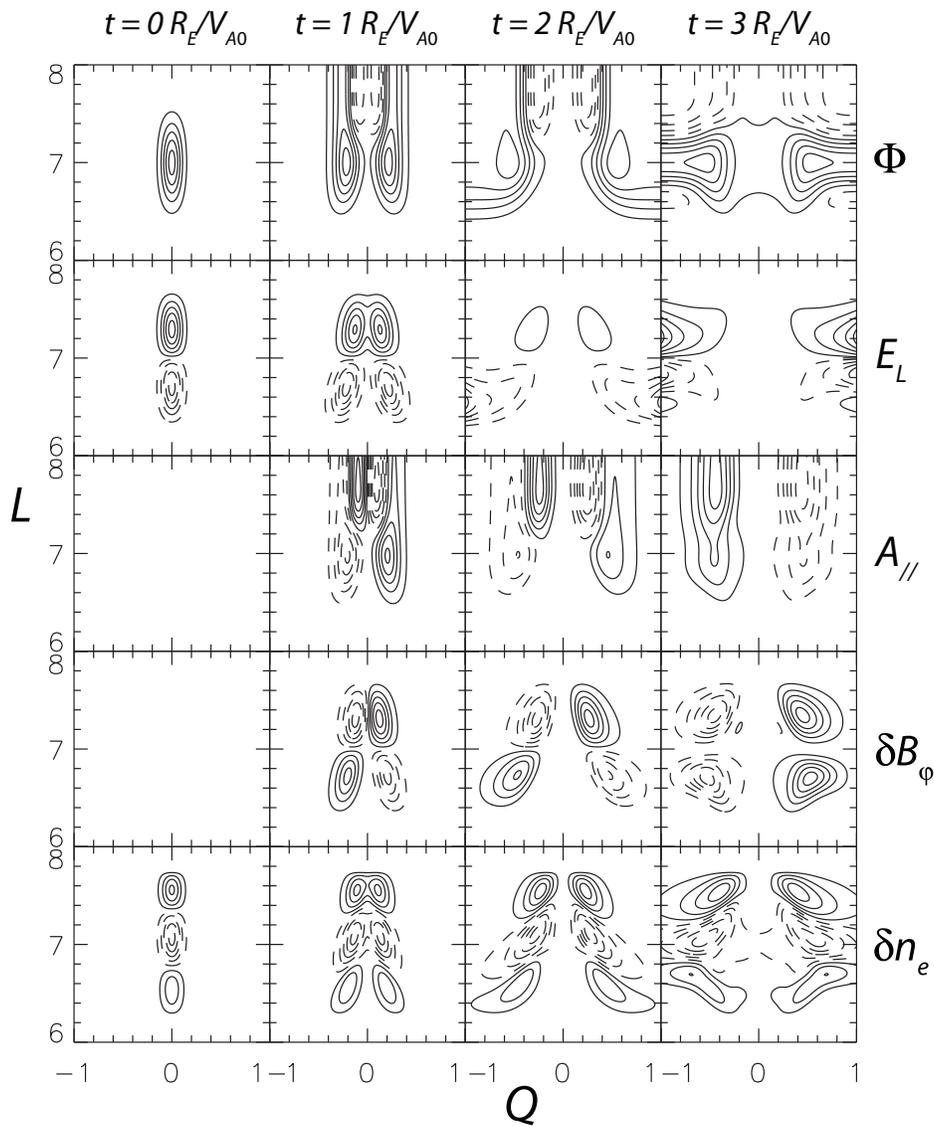} 
\caption{\label{F:LinSimResults} Contours of (from top to bottom) $\Phi$, 
$E_L$, $A_\|$, $\delta B_\varphi$, and $\delta n_e$ at four different 
times (arranged from left to right) indicated at the top of the plot.  In 
each panel, the contours are plotted versus the curvilinear coordinates $Q$ 
and $L$ and positive (negative) contour levels are solid (dashed).
}
\end{figure*}

Figure~\ref{F:LinSimResults} shows contours of $\Phi$, $E_L$, $A_\|$, $\delta B_\varphi$ (out of a plane component), and $\delta n_e$ with respect to the 
curvilinear coordinates $Q$ and $L$.  These are plotted at four different times normalized to $R_E/V_{A0}$, where $V_{A0}$ is the \Alfven\ 
speed at $L = 7$ and at the magnetic equator. The initial perturbation in $\Phi$ is plotted in the top left panel ($t = 0$).  The shape of the contours 
is different from those of Fig.~\ref{F:DipoleGrid} because of the different coordinate system, but it is clear from both figures that we have a single monotonic peak.  In addition to the initial perturbation in $\Phi$, there is also an initial perturbation in $n_e$ (bottom left panel).  Not shown are the initial perturbations for $n_i$, $p_{\| i}$, $p_{\perp i}$, $p_{\| e}$, and $p_{\perp e}$.  (There is no initial magnetic field perturbation.) All the variables are initialized with parameters consistent with a linear wave with parallel and perpendicular wavelengths corresponding to the scale lengths of the initial perturbation in the parallel and perpendicular directions.  

As can be seen from Fig.~\ref{F:LinSimResults}, the initial perturbation breaks up into two traveling waves that move along the magnetic field (left or right).  The traveling waves have a magnetic perturbation.  The \Alfven\ speed $\sim B_0$ and $B_0 \sim L^{-3}$.  Because of this, the perturbation travels faster at low values of $L$.  In addition, the real length of the field line is less at lower $L$ (Fig.~\ref{F:DipoleGrid}).   Therefore the perturbation travels much faster along the curvilinear coordinate $Q$ at the low values of $L$.  There are two lobes of the perturbation $E_L$ or 
$\delta B_\varphi$.  By the final time of the simulation, the lobes at lower $L \sim 6.5$ have already reflected off of the insulating boundary at $Q = \pm 1$, while the lobes at higher $L \sim 7.5$ have not.  In the MHD limit, one would associate the density perturbation with a sound wave rather than an \Alfven\ wave, but Fig.~\ref{F:LinSimResults} shows that all the perturbations travel together, consistent with a coherent linear wave.

\subsection{Nonlinear simulation in dipole geometry}
\label{S:nonlinSim}

\begin{figure}
\includegraphics[width=15pc]{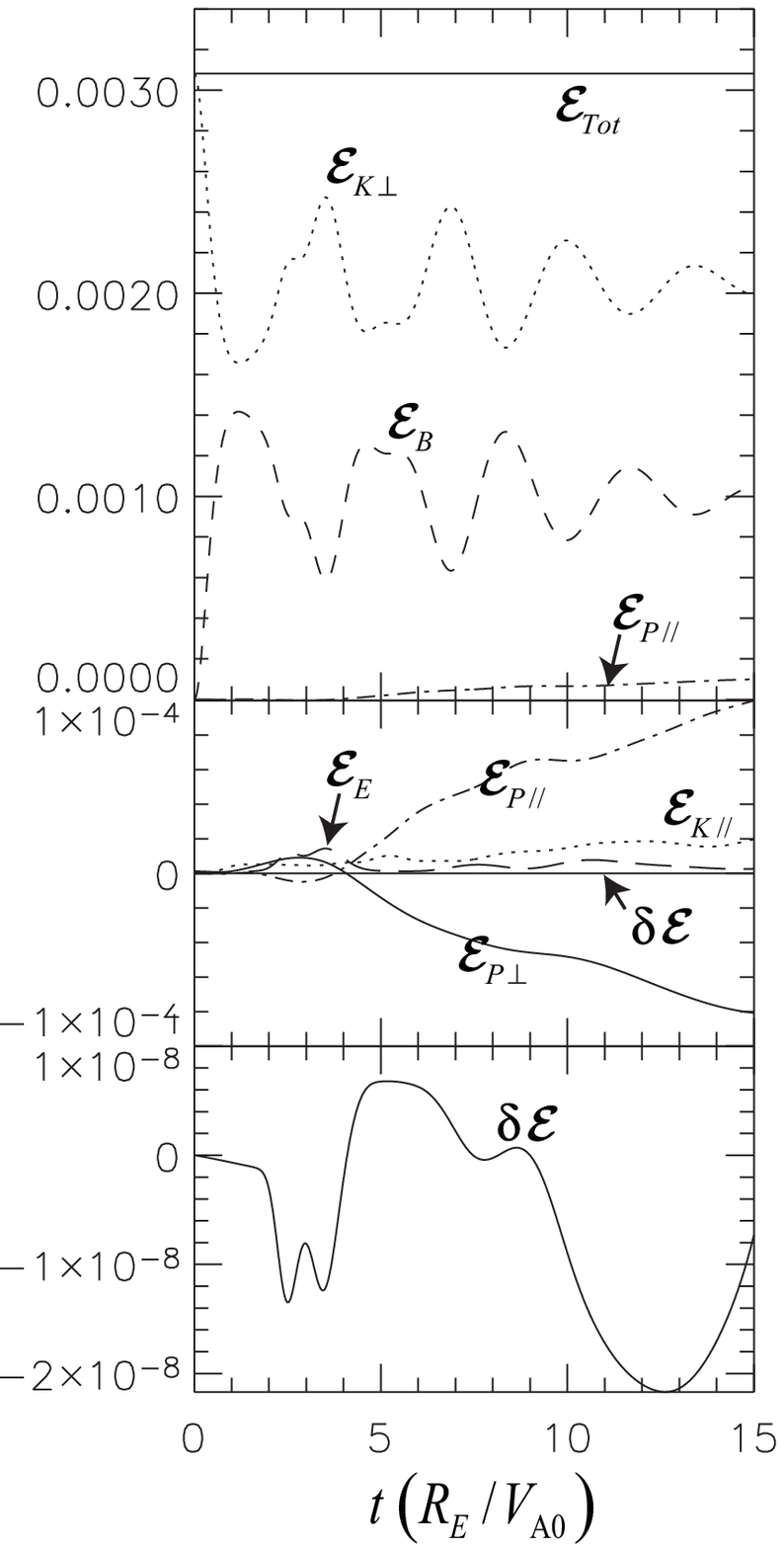} 
\caption{\label{F:NonlinEnCon} Energy terms versus time for   
  a nonlinear ($\delta v_\varphi / V_{A0} = 0.3$) simulation in dipole  
  geometry with parameters given in Table \ref{tab:sim_par}.  We plot in (a), the largest energy 
  terms, in (b), the smaller energy terms with the change in the total energy 
  $\delta {\cal E}$, and in (c), just $\delta {\cal E}$.
}
\end{figure}

Finally, we ran a third simulation with the same parameters as for the linear run in dipole geometry (Fig.~\ref{F:NonlinEnCon}), but with initial velocity perturbation $\delta v_\varphi / V_A = 0.3$.  This amplitude is quite large, and when a simulation is run in straight geometry with a purely sinusoidal perturbation, the fluctuations of the energy terms are not regular as they are in Fig.~\ref{F:LinEnCon} (not shown; especially irregular are the fluctuations in the parallel kinetic and pressure energies).  Figure~\ref{F:NonlinEnCon} (similar to Fig.~\ref{F:LinEnCon}) displays the energy terms for the nonlinear dipole simulation, and demonstrates that the total simulation energy is well conserved in this case also.

To demonstrate the convergence properties of this nonlinear energy conservation, we now focus our attention on the simulation time interval ranging from $t\,(V_{A0} / R_E) =$ 0 to 1 in Fig.~\ref{F:NonlinEnCon}, where the largest changes in energy are for the magnetic energy and the perpendicular kinetic energy. Figure \ref{F:convergence} shows the log-log plot of the error in the total energy $\delta{\cal E}$ normalized to the magnetic energy ${\cal E}_B$ (which is approximately the same for all the simulation results) as a function of the normalized time step $\Delta t \,(V_{A0} / R_E)$ for different grid resolutions $N_{i} = 128, 256,$ and 512 (the same in both the field-line $Q$-direction and radial $L$-direction). In the best-case scenario, we should see the following behavior as we reduce $\Delta t$ at a fixed value of $N_{i}$: Since the time step algorithm is a second-order predictor-corrector scheme (leapfrog trapezoidal), the error in the energy should go down as a factor of four for every reduction in $\Delta t$ of a factor of two. The error in the energy should then converge to a constant value limited by the spatial resolution (as observed in 
Fig.~\ref{F:convergence}); note that if we kept decreasing $\Delta t$, the error in the energy would eventually rise because the computer calculations would not have the precision necessary to accurately solve the equations. Because the spatial-differencing scheme for our simulation code is spatially fourth-order accurate, we should see a decrease in the energy error of a factor of $2^4 = 16$ each time we double the resolution (for a fixed time step $\Delta t$), and we do in fact see that decrease in the time-resolved error in Fig.~\ref{F:convergence} as we increase the number of grid points $N_{i}$ from 128 to 256 (decrease in error of $2.1 \times 10^{-6} / 1.3 \times 10^{-7} \simeq 16$) and from $N_{i}$ = 256 to 512 (decrease in error of $1.3 \times 10^{-7} / 8.0 \times 
10^{-9} \simeq 16$).

\begin{widetext}

\begin{figure}
\includegraphics[width=30pc]{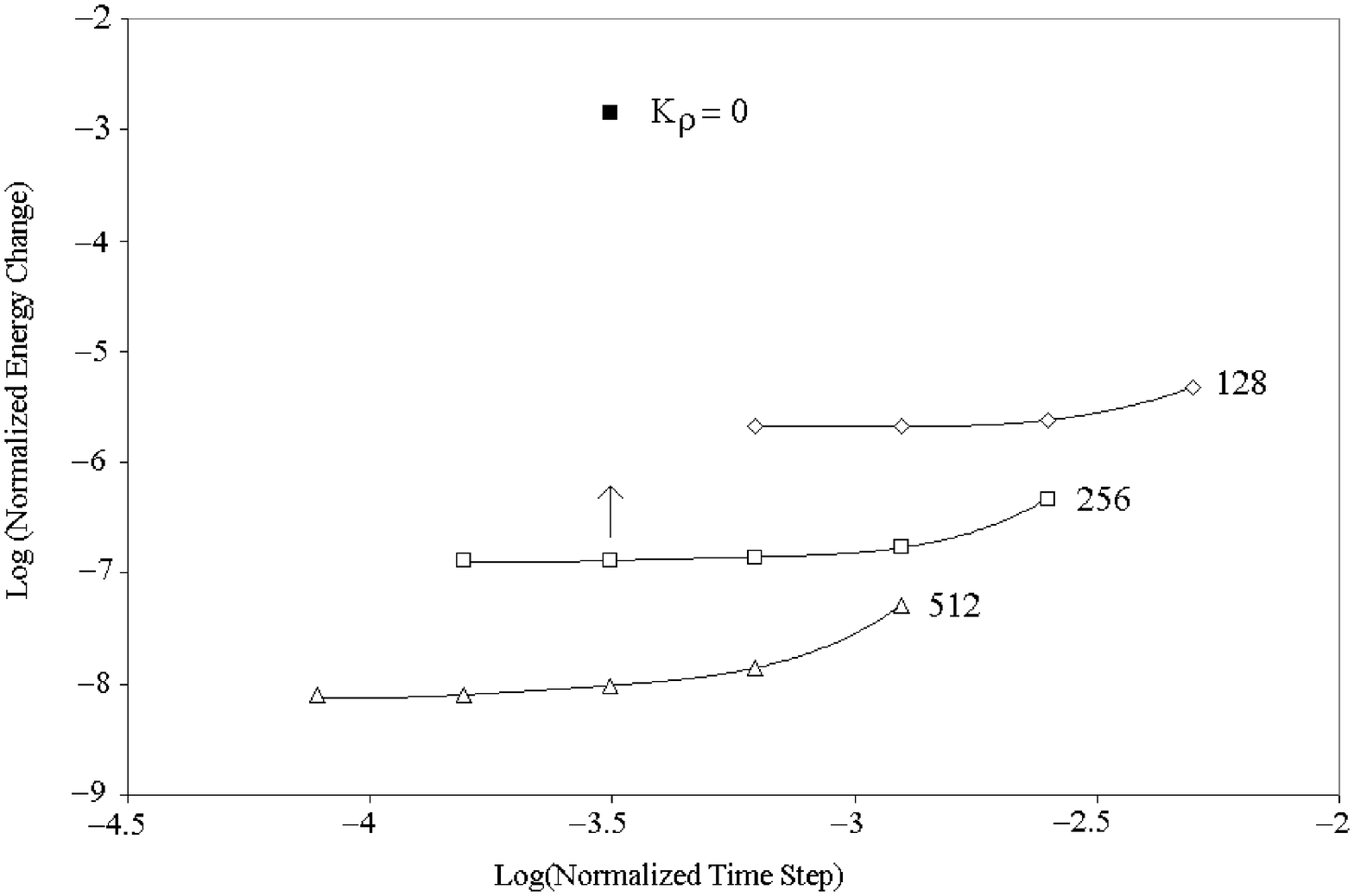} 
\caption{\label{F:convergence} For the nonlinear simulation in dipole geometry (for the time interval $t\,(V_{A0} / R_E) =$ 0 to 1 in 
Fig.~\ref{F:NonlinEnCon}), the logarithm of the normalized total energy $\delta{\cal E}/{\cal E}_B$ is shown as a function of the logarithm of the normalized time step $\Delta t\,(V_{A0}/ R_E)$ for different numbers of grid points $N_{i} = 128, 256, 512$ (the same number in both directions). For a fixed number of grid points $N_{i}$ (lines are used as guides), each successive normalized time step is reduced by half as we proceed to the left. In addition, the impact of setting $K_{\rho}$ in Eq.~(\ref{eq:F_rho}) equal to zero (that is, setting the nonlinear ponderomotive force density equal to zero) is reflected by a large jump (four orders of magnitude) in the total-energy non-conservation.}
\end{figure}

\end{widetext}

While we regard a detailed description of the physics of the new FLR terms as beyond the scope of this paper, we can easily demonstrate that they have an appreciable effect on the energy conservation.  If we run a simulation with $N_{i}$ = 256 and $\Delta t\,(V_{A0} / R_E) = 3.125 \times 10^{-4}$ (converged with respect to time for this grid resolution), but setting $K_{\rho}$ in Eq.~(\ref{eq:F_rho}) equal to zero (that is, setting the nonlinear ponderomotive force density equal to zero), we find the normalized energy error to be $\delta{\cal E}/{\cal E}_B = 1.4 \times 10^{-3}$, which is much larger than the value $1.3 \times 10^{-7}$ shown in Fig.~\ref{F:convergence}.  The fact that this error is less than unity indicates that the zeroth order physics of \Alfven\ waves (energy transfer between magnetic and perpendicular kinetic energy) is being correctly described.  However, the ratio of the change of the total energy is 43 times greater than that of the electric field energy and 1200 times greater than the parallel kinetic energy, showing that the parallel dynamics and dynamics associated with the displacement current are not at all well described.

\section{\label{sec:summary}Summary}

The variational derivation of the reduced fluid equations has revealed the existence of a new type on nonlinearity in reduced fluid dynamics. While standard fluid nonlinearities appear in the convective derivative operator ${\bf u}\bdot\nabla$ (e.g., $E\times B$ and magnetic-flutter nonlinearities), the new nonlinear terms presented here can be described as nonlinear FLR effects that appear as corrections (\ref{eq:PhiA_rho}) of the electromagnetic potentials $\Phi$ and $A_{\|}$. These nonlinear FLR-corrected potentials are then used to construct the magnetic and electric fields (\ref{eq:Bbot_rho}) and (\ref{eq:E_rho}) that appear in the parallel reduced-fluid equation (\ref{eq:uparallel_rho}), which also contains a ponderomotive-force correction 
(\ref{eq:CGL_rho}) to the standard CGL-pressure force density. 

The linear properties of the equations for a two-fluid homogeneous plasma were described, and linear and non-linear simulations demonstrated that the equations describe the coupled dynamics of \Alfven\ and sound waves and that the simulation energy is conserved in both straight and dipole geometry.

Lastly, we note that the single limitation on the background magnetic field in the present work was associated with the absence of parallel current along the field lines (i.e., $\bhat_{0}\bdot\nabla\btimes\bhat_{0} = 0$). This constraint was used as a simplifying assumption [see Eqs.~(\ref{eq:Bstar_def})-(\ref{eq:bunit_star})] associated with the perturbed magnetic field $\nabla\btimes(A_{\|}\,\bhat_{0}) \equiv {\bf B}_{\bot}$ having no component along the background field lines ($\bhat_{0}\bdot{\bf B}_{\bot} \equiv 0$). More general magnetic geometries with $\bhat_{0}\bdot\nabla\btimes\bhat_{0} \neq 0$ (e.g., tokamak geometry) can also be treated within a variational formulation and will be the subject of future work.

\begin{acknowledgments}
  Work at Dartmouth was supported by NASA grants NNG05GJ70G (Heliophysics Theory   
  Program) and NNG04GE22G and by NSF grant ATM-0120950 
  [Center for Integrated Space Weather Modeling (CISM), funded 
  by the Science and Technology Centers Program].
\end{acknowledgments}

\appendix

\section{\label{sec:push}Push-forward representation of fluid moments}

The purpose of this Appendix is to present the theoretical foundations \cite{Brizard_2006} that establish the relation between fluid moments of a particle distribution function $F_{{\rm phys}}({\bf z},t)$ in particle phase space and fluid moments of a reduced distribution function $F({\bf Z},t)$ in reduced phase space. Here, the near-identity (reversible) transformation ${\cal T}_{\epsilon}: {\bf z} \rightarrow {\bf Z}$ from particle phase space to reduced phase space is represented in asymptotic form as 
\[ Z^{\alpha} \;=\; z^{\alpha} + \epsilon\,G_{1}^{\alpha} + \epsilon^{2} \left( G_{2}^{\alpha} + \frac{1}{2}\,
{\sf G}_{1}\cdot dG_{1}^{\alpha} \right) + \cdots, \]
where $\epsilon \ll 1$ is a small ordering parameter and $({\sf G}_{1}, {\sf G}_{2}, \cdots)$ represent the Lie-transform vector fields that generate the transformation ${\cal T}_{\epsilon}$. Furthermore, this transformation induces the pull-back operator ${\sf T}_{\epsilon}: F \rightarrow 
F_{{\rm phys}} = {\sf T}_{\epsilon}F$ and push-forward operator ${\sf T}_{\epsilon}^{-1}: F_{{\rm phys}} \rightarrow F = {\sf T}_{\epsilon}^{-1}
F_{{\rm phys}}$ between phase-space functions, which both preserve the scalar invariance property $F_{{\rm phys}}({\bf z},t) = F({\bf Z},t)$. 

The reduced displacement 
\begin{equation}
\vb{\rho}_{\epsilon} \;\equiv\; {\sf T}_{\epsilon}^{-1}{\bf x} \;-\; {\bf X}
\label{eq:rho_epsilon} 
\end{equation}
(e.g., gyroradius) between the reduced position (e.g., guiding-center position) and the push-forward of the particle position is of particular importance here and it is expressed as
\[ \vb{\rho}_{\epsilon} \;=\; -\,\epsilon\;G_{1}^{{\bf x}} \,-\, \epsilon^{2} \left( G_{2}^{{\bf x}} - \frac{1}{2}\,{\sf G}_{1}\cdot\exd G_{1}^{{\bf x}} \right) + \cdots \]
in terms of the Lie-transform generating vector fields $({\sf G}_{1}, {\sf G}_{2}, \cdots)$.

\subsection{General reduced fluid moments}

We begin by considering an arbitrary fluid moment on physical (phys) particle phase space:
\begin{eqnarray}
(n\,[\chi])_{{\rm phys}} & = & \int d^{3}p\;\chi\;F_{{\rm phys}} \nonumber \\
 & = & \int d^{6}z\;\chi\,\delta^{3}({\bf x} - {\bf r})\;F_{{\rm phys}} \nonumber \\
 & = & \int d^{6}Z\; {\sf T}_{\epsilon}^{-1}\chi\;\delta^{3}({\bf X} + \vb{\rho}_{\epsilon} - {\bf r})\;F, 
\label{chi_def}
\end{eqnarray}
where $\chi$ is an arbitrary function on particle phase space and ${\sf T}_{\epsilon}^{-1}\chi$ is its push-forward on reduced phase space. Here, 
$[\chi]_{{\rm phys}}$ denotes the physical particle-momentum average of $\chi$ with respect to $F_{{\rm phys}}$ and $n_{{\rm phys}}$ is the particle fluid density. Upon Taylor expanding Eq.~(\ref{chi_def}) in powers of $\vb{\rho}_{\epsilon}$ and integrating by parts, we find the nonlinear FLR 
expansion
\begin{eqnarray} 
(n\,[\chi])_{{\rm phys}} & = & n\;\left[{\sf T}_{\epsilon}^{-1}\chi\right] \;-\; \nabla\bdot \left( n\;\left[
\vb{\rho}_{\epsilon}\;{\sf T}_{\epsilon}^{-1}\chi\right]\right) \nonumber \\
 &  &+\; \cdots 
\label{eq:chi_push}
\end{eqnarray}
where $[\cdots]$ denotes the reduced-momentum average with respect to $F$ and $n$ denotes the reduced fluid density. Note that the push-forward representation (\ref{eq:chi_push}) is reversible and its inverse yields the pull-back representation \cite{Brizard_1992}
\begin{eqnarray} 
n\,[\chi] & = & n_{{\rm phys}}\;\left[{\sf T}_{\epsilon}\chi\right]_{{\rm phys}} \;+\; \nabla\bdot \left( n_{{\rm phys}}\;\left[
\vb{\rho}_{\epsilon}\;{\sf T}_{\epsilon}\chi\right]_{{\rm phys}} \frac{}{} \right) \nonumber \\
 &  &+\; \cdots,
\label{eq:chi_pull}
\end{eqnarray}
where $\chi$ is an arbitrary function on reduced phase space and ${\sf T}_{\epsilon}\chi$ is its pull-back on particle phase space. 

We now consider the fluid moments associated with fluid density $(\chi = 1)$ and fluid velocity $(\chi = {\bf v} = 
d{\bf x}/dt)$. First, the push-forward representation for the particle fluid density
\begin{equation}
n_{{\rm phys}} \;=\; n \;-\; \nabla\bdot\left( n\;[\vb{\rho}_{\epsilon}] \;-\; \frac{1}{2}\;\nabla\bdot(n\;[\vb{\rho}_{\epsilon}
\vb{\rho}_{\epsilon}]) \right),
\label{eq:n_push}
\end{equation}
where we have retained the second-order (``quadrupole'') term, which will prove useful in what follows \cite{Brizard_2006}. 

Next, using the definition (\ref{eq:rho_epsilon}), we consider the push-forward representation for the particle fluid velocity 
\begin{equation}
{\sf T}_{\epsilon}^{-1}{\bf v} \;=\; \left[ {\sf T}_{\epsilon}^{-1}\frac{d}{dt}{\sf T}_{\epsilon}
\right]\left({\sf T}_{\epsilon}^{-1}{\bf x}\right) \;\equiv\; \frac{d_{\epsilon}{\bf X}}{dt} \;+\; 
\frac{d_{\epsilon}\vb{\rho}_{\epsilon}}{dt},
\label{eq:v_def}
\end{equation}
where $d_{\epsilon}/dt$ denotes the reduced Vlasov operator so that $d_{\epsilon}{\bf X}/dt$ denotes the reduced particle velocity (e.g., guiding-center velocity) and $d_{\epsilon}\vb{\rho}_{\epsilon}/dt$ denotes the reduced ``displacement" velocity (which may include the gyroangle-dependent perpendicular velocity and the gyroangle-independent polarization velocity). The push-forward representation for the particle flux 
$\vb{\Gamma}_{{\rm phys}} \equiv (n\,[{\bf v}])_{{\rm phys}}$:
\begin{eqnarray}
\vb{\Gamma}_{{\rm phys}} & = & n\;\left[ \frac{d_{\epsilon}{\bf X}}{dt}\right] \;+\; \pd{}{t} \left( \frac{}{} n\;[\vb{\rho}_{\epsilon}]\right) 
\nonumber \\
 &  &+\; \nabla\btimes\left( n\;\left[\vb{\rho}_{\epsilon}\btimes \left( \frac{1}{2}\,\frac{d_{\epsilon}\vb{\rho}_{\epsilon}}{dt} \;+\; 
\frac{d_{\epsilon}{\bf X}}{dt} \right)\right] \right) \nonumber \\
 & \equiv & \vb{\Gamma} \;+\; \vb{\Gamma}_{{\rm pol}} \;+\; \vb{\Gamma}_{{\rm mag}}
\label{eq:v_push}
\end{eqnarray}
is expressed in terms of the reduced flux $\vb{\Gamma} = n\,[d_{\epsilon}{\bf X}/dt]$, the polarization flux
\begin{equation}
\vb{\Gamma}_{{\rm pol}} \;\equiv\; \pd{}{t} \left( \frac{}{} n\;[\vb{\rho}_{\epsilon}]\right),
\label{eq:pol_def}
\end{equation}
and the reduced (divergenceless) magnetization flux 
\begin{equation} 
\vb{\Gamma}_{{\rm mag}} \;\equiv\; \nabla\btimes\left( \frac{n}{2}\;\left[\vb{\rho}_{\epsilon}\btimes
\frac{d_{\epsilon}\vb{\rho}_{\epsilon}}{dt}\right] \;+\; n\;\left[\vb{\rho}_{\epsilon}\btimes
\frac{d_{\epsilon}{\bf X}}{dt}\right] \right),
\label{eq:mag_def}
\end{equation}
which is itself decomposed in terms of an intrinsic contribution (first term) and a moving-dipole contribution (second term) \cite{Jackson_1975}. Note that the correct derivation of the intrinsic contribution relied on keeping the quadrupole contribution in Eq.~(\ref{eq:n_push}).

The push-forward representations (\ref{eq:n_push}) and (\ref{eq:v_push}) preserve the conservation law of particles through the continuity equation
\begin{eqnarray*}
0 & = & \pd{n_{{\rm phys}}}{t} \;+\; \nabla\bdot\vb{\Gamma}_{{\rm phys}} \\
 & = & \pd{}{t} \left[n - \nabla\bdot\left( n\,[\vb{\rho}_{\epsilon}]\right) \frac{}{} \right] \;+\; \nabla\bdot\left( 
\vb{\Gamma} + \vb{\Gamma}_{{\rm pol}} + \vb{\Gamma}_{{\rm mag}} \right) \\
 & = & \pd{n}{t} \;+\; \nabla\bdot\vb{\Gamma},
\end{eqnarray*}
where the reduced polarization effects cancel each other exactly and the reduced magnetization flux is divergenceless (by definition).

Lastly, using these push-forward representations (\ref{eq:n_push}) and (\ref{eq:v_push}), we may write the push-forward representation for charge density as
\[ \rho_{{\rm phys}} \;\equiv\; \sum\;q\,n_{{\rm phys}} \;=\; \rho \;-\; \nabla\bdot{\bf P}_{\epsilon}, \]
where $\rho = \sum\,q\,n$ and the reduced polarization vector is
\[ {\bf P}_{\epsilon} \;\equiv\; \sum\;q\;n\;[\vb{\rho}_{\epsilon}] \;+\; \cdots, \]
and the push-forward representation for current density as
\[ {\bf J}_{{\rm phys}} \;\equiv\; \sum\;q\,\vb{\Gamma}_{{\rm phys}} \;=\; {\bf J} \;+\; \pd{{\bf P}_{\epsilon}}{t} \;+\; c\;
\nabla\btimes{\bf M}_{\epsilon}, \]
where ${\bf J} = \sum\,q\,\vb{\Gamma}$ and the reduced magnetization vector is
\[ {\bf M}_{\epsilon} \;=\; \sum\;\frac{qn}{c}\; \left[\vb{\rho}_{\epsilon}\btimes\left( \frac{1}{2}\;
\frac{d_{\epsilon}\vb{\rho}_{\epsilon}}{dt} \;+\;\frac{d_{\epsilon}{\bf X}}{dt} \right)\right]  \]
The reduced Maxwell equations are, thus, expressed in terms of the ``macroscopic'' fields ${\bf D} \equiv {\bf E} + 4\pi\,{\bf P}_{\epsilon}$ and 
${\bf H} \equiv {\bf B} - 4\pi\,{\bf M}_{\epsilon}$ as
\[ \nabla\bdot{\bf D} \;=\; 4\pi\,\rho \;\;{\rm and}\;\; \nabla\btimes{\bf H} \;-\; 
\frac{1}{c}\,\pd{{\bf D}}{t} \;=\; \frac{4\pi}{c}\,{\bf J}. \]

\subsection{Guiding-center and gyrocenter fluid moments}

Let us now obtain explicit expressions for the reduced fluid moments associated with the guiding-center and gyrocenter phase-space transformations. 

We begin with the guiding-center (gc) transformation for which $\vb{\rho}_{\epsilon} = \vb{\rho}_{0}$ (and we ignore electric and magnetic perturbation fields until we consider the next transformation). By substituting the gyroangle-dependent gyroradius $\vb{\rho}_{0}$ in Eq.~(\ref{eq:n_push}), we easily find that the physical and guiding-center densities are equal $n_{{\rm phys}} = n_{{\rm gc}}$, since we ignore standard FLR corrections in the present work; in the same spirit, the perpendicular and parallel pressures are also identical in the physical and guiding-center fluid descriptions. The relation (\ref{eq:v_push}) between the physical particle flux and the guiding-center particle flux, on the other hand, yields the expression
\begin{eqnarray}
\vb{\Gamma}_{{\rm phys}} & = & \vb{\Gamma}_{{\rm gc}} \;-\; \nabla\btimes\left( \frac{p_{\bot}\,\bhat}{m\,\Omega} \right) \nonumber \\
 & = & n\,u_{\|}\;\bhat \;+\; \frac{\bhat}{m\Omega}\btimes\nabla\bdot{\sf P},
\label{eq:Gamma_gc}
\end{eqnarray}
where the guiding-center flux is
\[ \vb{\Gamma}_{{\rm gc}} \;=\; n\,u_{\|}\;\bhat \;+\; \frac{\bhat}{m\Omega}\btimes \left( p_{\bot}\;\nabla\ln B \;+\; p_{\|}\;\bhat\bdot\nabla\bhat
\right), \] 
only the guiding-center parallel magnetization $(-\,p_{\bot}\bhat/m\Omega)$ survives gyroangle averaging, and $\nabla\bdot{\sf P}$ denotes the divergence of the CGL pressure tensor (\ref{eq:CGL_P}). Here, the parallel fluid velocities are also identical in the physical and guiding-center fluid descriptions (labeled $u_{\|}$). Note that this expression is equivalent to the reduced-fluid velocity (\ref{eq:u_vector}) in the absence of electric and magnetic perturbation fields.

Next, we consider the gyrocenter (gy) transformation (now involving the electric and magnetic perturbation fields) for which we take $[\vb{\rho}_{\epsilon}] = \vb{\rho}_{\bot}$. In this case, using Eq.~(\ref{eq:n_push}), the guiding-center (and physical) density is expressed 
\begin{equation}
n_{{\rm gc}} \;=\; n_{{\rm gy}} \;-\; \nabla\bdot\left( n_{{\rm gy}} \;\vb{\rho}_{\bot} \right).
\label{eq:n_gyro}
\end{equation}
The relation (\ref{eq:v_push}) between the physical particle flux and the guiding-center particle flux, on the other hand, yields the expression
\begin{eqnarray}
\vb{\Gamma}_{{\rm gc}} & = & \vb{\Gamma}_{{\rm gy}} \;+\; \pd{}{t} \left( n_{{\rm gy}}\;\vb{\rho}_{\bot} \right) \nonumber \\
 &  &+\; \nabla\btimes\left[(n\,u_{\|})_{{\rm gy}}\; \vb{\rho}_{\bot}\btimes\bhat_{0} \right],
\label{eq:Gamma_gy}
\end{eqnarray}
which includes the polarization-drift flux and the (moving-dipole) magnetization flux to first order in NFLR corrections. This expression can be used to derive the parallel flux relation
\begin{eqnarray}
\Gamma_{\|{\rm gc}} & = & (n\,u_{\|})_{{\rm gy}} \;+\; \bhat_{0}\bdot\nabla\btimes\left[(n\,u_{\|})_{{\rm gy}}\; \vb{\rho}_{\bot}\btimes\bhat_{0} 
\right], \nonumber \\
 & = & \left[ n_{{\rm gy}} \;-\; \nabla\bdot\left( n_{{\rm gy}} \;\vb{\rho}_{\bot} \right) \right] u_{\|{\rm gy}} \nonumber \\
 &  &-\; n_{{\rm gy}}\;\left(\vb{\rho}_{\bot}\bdot\nabla u_{\|{\rm gy}}\right) \;+\; \cdots,
\label{eq:Gamma_pargy}
\end{eqnarray}
where we have ignored the effects of background magnetic curvature in obtaining the second equality. By using the relation (\ref{eq:n_gyro}) between the guiding-center and gyrocenter densities, we now obtain the relation
\begin{equation}
u_{\|{\rm gc}} \;=\; u_{\|{\rm gy}} \;-\; \vb{\rho}_{\bot}\bdot\nabla u_{\|{\rm gy}} \;+\; \cdots,
\label{eq:u_gy}
\end{equation}
where we have ignored higher-order NFLR corrections.

Lastly, we combine the guiding-center and gyrocenter push-forward relations to obtain the expressions for the physical (particle) density
\begin{equation}
n_{{\rm phys}} \;=\; n \;-\; \nabla\bdot\left(n\;\vb{\rho}_{\bot}\right),
\label{eq:n_gcgy}
\end{equation}
and the physical (particle) flux
\begin{eqnarray}
\vb{\Gamma}_{{\rm phys}} & = & n\,{\bf u} \;+\; \pd{}{t} \left( n\;\vb{\rho}_{\bot} \right) \nonumber \\
 &  &+\; \nabla\btimes\left( n\; \vb{\rho}_{\bot}\btimes u_{\|}\,\bhat_{0} \right),
\label{eq:Gamma_gcgy}
\end{eqnarray}
in terms of the gyrocenter density $n$ and the gyrocenter velocity ${\bf u}$ (used in the text). 

\section{\label{sec:pol_disp}Reduced-fluid Polarization Displacement}

Since $\vb{\rho}_{\bot}$ is the key quantity describing our new interpretation of the reduced fluid equations (first presented in Ref.~\cite{Brizard_2005}), it is important to have a good understanding of what it is.  The displacement vector $\vb{\rho}_{\bot}$ is shown here to possess a simple interpretation as the time integrated inertial drift \citep{Schmidt}.  

Neglecting curvature, we can easily derive a formula for $\vb{\rho}_{\bot}$ from the single particle equation of motion.  Writing the particle velocity as 
\begin{subequations}
  \label{E:simple_vExpansion}
  \begin{eqnarray}
    {\bf v} & = & {\bf v}_0 + {\bf v}_{1\bot}, \\
    {\bf v}_0 & = & u_\| \left( \bhat_{0} + \frac{{\bf B}_{\bot}}{B_0} 
       \right) + {\bf u}_{E},
  \end{eqnarray}
\end{subequations}
we approximate the equation of motion as
\begin{eqnarray}
  \label{E:simple_equationOfMotion}
  \frac{d{\bf v}_0}{dt} & \simeq & \frac{q}{m} \left( {\bf E} + 
     \frac{1}{c} {\bf v}\btimes{\bf B} \right) \nonumber \\
 & = & \frac{q}{mc}\;{\bf v}_{1\bot}\btimes{\bf B}_{0}.
\end{eqnarray}
Solving for ${\bf v}_{1\bot}$, we find
\begin{eqnarray}
  \label{E:simple_v_one}
  {\bf v}_{1\bot} & = & \frac{\bhat_{0}}{\Omega_0} \btimes \frac{d{\bf v}_0}{dt} 
      \nonumber \\
    & = & \frac{d}{dt} \left( \frac{1}{\Omega_0} \bhat_{0} \btimes {\bf v}_0 \right) 
      \nonumber \\
    & = & \frac{d}{dt} \left(\frac{\bhat_{0}}{\Omega_{0}}\btimes\left( {\bf u}_{E} \;+\; u_{\|}\;
\frac{{\bf B}_{\bot}}{B_{0}} \right) \right),
\end{eqnarray}
where we neglected curvature effects in going from the first line to the second line. Identifying 
\begin{equation}
{\bf v}_{1\bot} \;=\; d\vb{\rho}_{\bot}/dt 
\label{eq:v1_rho}
\end{equation}
immediately leads to Eq.~(\ref{eq:df_dipole}).

Next, we now present a simple derivation of the NFLR-corrected potentials (\ref{eq:PhiA_rho}) based on the electromagnetic one-form ${\sf A} \equiv 
{\bf A}\bdot\exd{\bf x} - \Phi\,c\,\exd t$. By expanding the NFLR-corrected one-form 
\[ {\sf A}_{\rho} \;\equiv\; {\bf A}({\bf x} + \vb{\rho}_{\bot},t)\bdot (\exd{\bf x} + \exd\vb{\rho}_{\bot}) \;-\; \Phi({\bf x} + \vb{\rho}_{\bot}, t)\;
c\,\exd t \]
to first order in $\vb{\rho}_{\bot}$, we obtain
\begin{eqnarray}
{\sf A}_{\rho} & = & {\sf A} \;+\; \vb{\rho}_{\bot}\bdot\left( \nabla{\bf A}\bdot\exd{\bf x} \;-\; \nabla\Phi\;c\,\exd t\right) \;+\; {\bf A}\bdot
\exd\vb{\rho}_{\bot} \nonumber \\
 & = & \left( {\bf A} \;-\; \vb{\rho}_{\bot}\btimes{\bf B}\right)\bdot \exd{\bf x} \;-\; \left( \Phi \;-\; \vb{\rho}_{\bot}\bdot{\bf E} \right)\;
c\,\exd t \nonumber \\
 &  &+\; \exd({\bf A}\bdot\vb{\rho}_{\bot}) \nonumber \\
 & \equiv & \left( {\bf A}_{\rho} \;+\; \nabla\eta \right)\bdot\exd{\bf x} \;-\; \left( \Phi_{\rho} \;-\; \frac{1}{c}\,\pd{\eta}{t} \right)\;
c\,\exd t,
\label{eq:A1_rho}
\end{eqnarray}
where 
\begin{equation}
\left( \begin{array}{c}
\Phi_{\rho} \\
\\
{\bf A}_{\rho}
\end{array} \right) \;=\; \left( \begin{array}{c}
\Phi \;-\; \vb{\rho}_{\bot}\bdot{\bf E} \\
\\
{\bf A} \;-\; \vb{\rho}_{\bot}\btimes{\bf B}
\end{array} \right),
\label{eq:phiA_def}
\end{equation}
and $\eta \equiv {\bf A}\bdot\vb{\rho}_{\bot}$ is treated as a gauge term. In Eq.~(\ref{eq:PhiA_rho}), the vector potential is ${\bf A} = A_{\|}\,
\bhat_{0}$ (i.e., $\eta \equiv 0$) and $A_{\|\rho} \equiv \bhat_{0}\bdot{\bf A}_{\rho}$.


\begin{thebibliography}{99}
\bibitem{footnote_1} For studies of \Alfven\ dynamics on auroral field lines \cite{DentonEA07}, we estimate that this parameter is large at or below one Earth radius altitude.  This so-called ``gap region'' (gap between the magnetosphere modelled by MHD codes and the 
ionospheric region) is of great interest for magnetosphere-ionosphere coupling.
\end{thebibliography}
\end{document}